\documentclass[aps,prx,reprint,groupedaddress]{revtex4-1}

\usepackage{graphicx}
\usepackage{layouts}

\newcommand{\com}{COM}

\usepackage{color}
\newcommand{\rev}[1]{#1}

\begin{document}

%\preprint{}

\title{Measuring the Internal Temperature 
%Measurement 
of a Levitated Nanoparticle in High Vacuum}



\author{Erik Hebestreit}
%\homepage[]{http://www.photonics.ethz.ch}
%\thanks{}
\affiliation{Photonics Laboratory, ETH Z\"{u}rich, 8093 Z\"{u}rich, Switzerland}

\author{Ren\'{e} Reimann}
\affiliation{Photonics Laboratory, ETH Z\"{u}rich, 8093 Z\"{u}rich, Switzerland}

\author{Martin Frimmer}
\affiliation{Photonics Laboratory, ETH Z\"{u}rich, 8093 Z\"{u}rich, Switzerland}

\author{Lukas Novotny}
\affiliation{Photonics Laboratory, ETH Z\"{u}rich, 8093 Z\"{u}rich, Switzerland}


\date{\today}

\begin{abstract}
The interaction of an object with its surrounding bath can lead to a coupling between the object's internal degrees of freedom and its center-of-mass motion. This coupling is especially important for nanomechanical oscillators, which are amongst the most promising systems for preparing macroscopic objects in quantum mechanical states.  Here we exploit this coupling to derive the {\em internal} temperature of a levitated nanoparticle from measurements of its center-of-mass dynamics.  For a laser-trapped silica particle in high vacuum we find an internal temperature of $1000(60)\,\mathrm{K}$.  The measurement and control of the internal temperature of nanomechanical oscillators is of fundamental importance because blackbody emission sets limits to the coherence of macroscopic quantum states. 
 %Nanomechanical oscillators, such as nanoparticles optically levitated in vacuum, are amongst the most promising systems for preparing macroscopic objects in quantum mechanical states. A great challenge for the realization of these quantum states is their decoherence due to blackbody emission. Therefore, measuring and controlling the internal temperature of these oscillators is of fundamental importance, especially for optically levitated objects, which suffer from absorption heating through the trapping light. We experimentally explore the coupling between the internal temperature of a levitated nanoparticle and its center-of-mass motion in high vacuum. We exploit this coupling, mediated through residual gas in the vacuum, to measure the internal temperature of a levitated particle, which is a crucial parameter in the context of future quantum protocols using levitated nanoparticles. We measure an internal temperature of our levitated particle in high vacuum of $1000(60)\,\mathrm{K}$.\\
  \end{abstract}

%\keywords{}

\maketitle

\section{Introduction}
Nanomechanical oscillators are used for inertial sensing and for the ultrasensitive measurement of forces and masses \cite{Degen2017,Mamin2001,Teufel2008,Jensen2008,Anetsberger2010,Moser2013}. Especially impressive are the latest advances to use them for producing and sensing mechanical quantum states \cite{Chan2011}. Amongst these oscillators, nanoparticles that are levitated in a strongly focused laser beam in vacuum stand out by lacking a mechanical clamping mechanism \cite{Li2011,Gieseler2012}, which makes them promising candidates for quantum interference experiments with macroscopic objects \cite{Chang2010,Romero-Isart2011,Arndt2014}. Recently, the center-of-mass (\com{}) temperature $T_\mathrm{com}$ of these levitated optomechanical systems has been reduced to only a few phonons using parametric \cite{Li2011, Gieseler2012, Jain2016}, or cavity-based cooling schemes \cite{Kiesel2013a,Asenbaum2013,Millen2015}, and is therefore quickly approaching the quantum regime.

Quantum states of such mesoscopic mechanical objects are extremely sensitive to decoherence, arising from any interaction of the object with its environment. For example, at finite pressure,  random collisions with residual gas molecules decohere the mechanical motion. In ultra-high vacuum, the radiation-pressure shot noise of the trapping laser is the dominant fluctuating force \cite{Jain2016}. Importantly, even in the absence of a trapping laser, for example during a matter-wave-interference free-fall experiment \cite{Hackermuller2004,Romero-Isart2011,Bateman2014}, there is a decoherence mechanism at work. This mechanism is the recoil force stemming from emitted blackbody photons due to the finite internal temperature of the trapped particle. In high-vacuum, where thermal emission is the only cooling channel for the internal particle temperature, even a minute absorption of the trapping laser by the particle can lead to an elevated internal temperature of a levitated object \cite{Chang2010,Romero-Isart2011,Bateman2014,Millen2014,Ranjit2015a,Rahman2015a,Juan2016,Rahman2017}. With the internal temperature limiting the lifetime of mesoscopic mechanical quantum states, the measurement and control of this internal temperature is a crucial prerequisite for future quantum interference experiments. 

%Quantum states, however, are known to be sensitive to decoherence through blackbody emission \cite{Hackermuller2004}, and the absorption of the trapping laser by the particle is understood to cause a high internal temperature $T_\mathrm{int}$ of vacuum levitated objects \cite{Chang2010,Romero-Isart2011,Bateman2014,Millen2014,Ranjit2015a,Rahman2015a,Juan2016,Rahman2017}. Accordingly, the measurement and control of the internal particle temperature are crucial, as it will limit the lifetime of future quantum states generated with levitated nanoparticles.

Moreover, even for levitated sensors in the classical regime, the internal temperature plays a central role. For instance, high internal temperatures are suspected to be the cause of particle loss at reduced gas pressures in numerous implementations \cite{Millen2014,Ranjit2015a,Rahman2015a}. Furthermore, a coupling between the internal temperature and the \com{} motion has become apparent in earlier studies of levitated particles at pressures above $1\,\mathrm{mbar}$ \cite{Millen2014}. This coupling leads to increased \com{} temperatures, which implies a reduced force sensitivity of  levitated sensors \cite{Ranjit2015a,Ranjit2016}. Despite its evident importance  in the field of levitated optomechanics, the internal temperature and its influence on the \com{} motion has not yet been studied in the high vacuum regime below $10^{-3}\,\mathrm{mbar}$, where the exceptional quality factors of levitated systems can be exploited for ultra-sensitive measurements, cooling of the \com{} temperature, and the future creation of mechanical quantum states. 

\rev{In this work, we measure the internal particle temperature in high vacuum ($10^{-3}$ to $10^{-6}\,\mathrm{mbar}$) by selectively heating the particle with an infrared laser that is much lower in power than the trapping laser. Accordingly, this separate heating laser allows us to vary the internal particle temperature without affecting the trapping potential.}
%, which keeps the optical trap unchanged. 
By measuring the thermal relaxation dynamics of the particle's \com{} motion at varying internal temperatures, we quantify the influence of the particle's internal temperature on its \com{} dynamics. We model the interactions of an internally hot particle with a dilute gas \cite{Millen2014} to determine the internal particle temperature and the energy accommodation coefficient, which characterizes the energy transfer between the particle's internal degrees of freedom and the surrounding gas. %Our measurements confirm that the internal temperature of current optically levitated nanoparticle systems limits the \com{} coherence and therefore quantum mechanical protocols in high vacuum.

\section{Experimental System}

%Our levitation system is realized in high vacuum and allows measuring and controlling the \com{} position of the optically trapped nanoparticle. For absolute measurements of the internal particle temperature $T_\mathrm{int}$, and to explore the coupling between the internal particle temperature and the \com{} temperature, we need to be able to quantitatively modify $T_\mathrm{int}$ without changing the optical trap. We achieve this through absorptive heating of the nanoparticle with a separate heating laser.

%\subsection{Optical Levitation System}

\begin{figure}
	\includegraphics{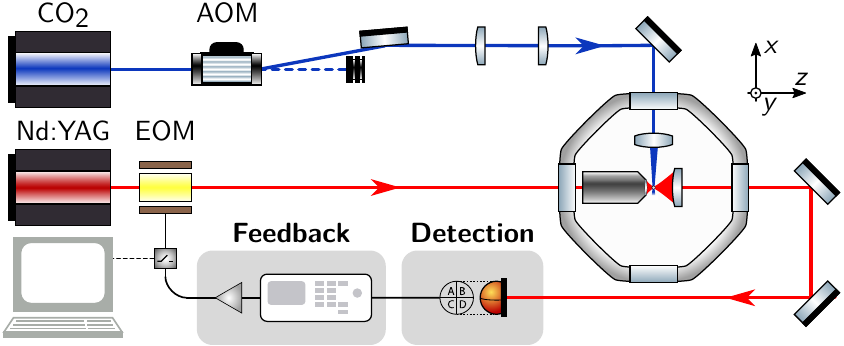}%
	\caption{Experimental setup. In the focus of a $1064\,\mathrm{nm}$ laser (red), we trap a $\mathrm{SiO_2}$ nanoparticle. The trap is placed in a vacuum chamber to adjust the gas pressure. The light scattered by the particle is collected and detected with a balanced detector. A switchable parametric feedback allows \rev{us to cool} the particle motion by modulating the laser beam \rev{intensity} with an electro-optic modulator (EOM). To increase the internal temperature of the particle, we irradiate it from the side with a $\mathrm{CO_2}$ laser (blue) whose intensity we adjust with an acousto-optic modulator (AOM). \label{fig:heating_setup}}
\end{figure}

Figure \ref{fig:heating_setup} shows our experimental setup. We use a Nd:YAG laser operating at a wavelength of $1064\,\mathrm{nm}$ with a power of $70\,\mathrm{mW}$ that we focus with an objective (numerical aperture $0.8$) to optically trap a $\mathrm{SiO_2}$ nanoparticle with a nominal radius of $R=68\,\mathrm{nm}$ inside a vacuum chamber. %The optical peak intensity at the particle position is $110\,\mathrm{mW/\mu m^2}$.
\rev{All data shown in this work is recorded using the same nanoparticle. Besides small variations in mass from particle to particle, we find identical results when using different particles.}
In forward direction, the scattered light is collected with an aspheric lens to detect the particle position along three axes ($x, y, z$) with a balanced detection scheme \cite{Gieseler2012}.
For small oscillation amplitudes, the \com{} motion of the levitated nanoparticle along the coordinate $x$ (equivalently for $y$ and $z$) follows the equation of motion of a damped harmonic oscillator
\begin{equation}\label{eq:eq-of-motion}
m\ddot{x}+m\gamma\dot{x}+m\Omega_0^2x=F_\mathrm{fluct}(t)\, .
\end{equation}
Here, $m$ is the particle mass, $\gamma$ the damping rate and $\Omega_0$ the oscillation frequency.
%, and the first and second derivative of the particle position $\dot{x}$ and $\ddot{x}$, respectively. 
Typical mechanical oscillation frequencies $\Omega_0/2\pi$ are $125\,\mathrm{kHz}$, $150\,\mathrm{kHz}$, and $40\,\mathrm{kHz}$ for the oscillation in $x$, $y$, and $z$ direction, respectively. The random fluctuating force $F_\mathrm{fluct}(t)$ describes the particle's interactions with the environment. Throughout this work, we operate in a pressure range of $10^{-3}$ to $10^{-6}\,\mathrm{mbar}$, where these interactions are dominated by collisions with gas molecules. In thermal equilibrium, the particle's \com{} motion is characterized by the temperature $T_\mathrm{com}$, which equals the bath temperature of the gas $T_\mathrm{bath}$. The \com{} temperature is given by $T_\mathrm{com}=m\left\langle \dot{x}^2\right\rangle/k_\mathrm{B}$, where $\left\langle \dot{x}^2\right\rangle$ is the variance of the particle velocity and $k_\mathrm{B}$ is the Boltzmann constant \cite{Hebestreit2017}. \rev{Using this relation between bath temperature and oscillation amplitude, we calibrate our detector signal at a pressure of $10\,\mathrm{mbar}$, where the gas temperature can be considered in equilibrium with room temperature. Subsequently, we transfer this calibration to low pressure by measuring the particle's response to an electric field as detailed in Ref.~\cite{Hebestreit2017}.}
%, which result in a thermal state of the \com{} motion. This thermal state is characterized by the \com{} temperature $T_\mathrm{com}=m\left\langle \dot{x}^2\right\rangle/k_\mathrm{B}$ \cite{Hebestreit2017}. The balance between fluctuating force and damping causes $T_\mathrm{com}$ to approach a steady state with the bath temperature $T_\mathrm{bath}$. These damping and heating processes originate from collisions with gas molecules, and, in ultra-high vacuum, from interactions with scattering photons \cite{Jain2016}. In the pressure range of $10^{-3}$ to $10^{-6}\,\mathrm{mbar}$, where we operate in this work, only the former is relevant, which is why in the following we solely concentrate on interactions of the particle with the surrounding gas. 
Using the position information obtained from the light scattered by the particle, we apply a parametric feedback in order to cool the particle's \com{} temperature to $T_\mathrm{com}\ll T_\mathrm{bath}$ \cite{Gieseler2012}. Table~\ref{tab:parameters} gives an overview of relevant parameters used in this work.

%\subsection{Modifying the Internal Temperature}

To control the internal temperature $T_\mathrm{int}$ of the optically levitated nanoparticle without changing the optical trapping potential, we use a $\mathrm{CO_2}$ laser with a wavelength of $10\,\mathrm{\mu m}$ (blue beam in Fig.~\ref{fig:heating_setup}) that illuminates the particle perpendicular to the propagation direction of the trapping beam. We adjust the $\mathrm{CO_2}$ laser intensity $I_\mathrm{CO_2}$ using an acousto-optic modulator (AOM). At a wavelength of $10\,\mathrm{\mu m}$ the optical absorption of silica is more than 7 orders of magnitude larger than at $1064\,\mathrm{nm}$~\cite{Kitamura2007}. Therefore, a weakly focused $\mathrm{CO_2}$ laser beam (numerical aperture $0.06$) with a peak intensity $I_\mathrm{CO_2}<1\,\mathrm{\mu W/\mu m^2}$ is sufficient to tune the particle's internal temperature for our experiments. Importantly, the optical forces generated by the $\mathrm{CO_2}$ laser on the particle are negligible. 
%At a wavelength of $10\,\mathrm{\mu m}$ silica has an absorption index of $\mathrm{Im}(n)\approx 1$, where $n$ is the refractive index. This exceeds the absorption at a wavelength of $1064\,\mathrm{nm}$ by 7 orders of magnitude \cite{Kitamura2007}. Therefore, only a weakly focused beam (numerical aperture $0.06$) with a peak intensity $I_\mathrm{CO_2}$ of less than $1\,\mathrm{\mu W/\mu m^2}$ is required to heat the particle's internal temperature sufficiently. This way, we can ensure that the trapping potential is not significantly disturbed by the heating laser. 
We estimate that the stiffening of the trapping potential due to the $\mathrm{CO_2}$ laser causes a shift of the oscillation frequency of less than $0.1\,\mathrm{mHz}$ and the radiation pressure of the $\mathrm{CO_2}$ laser results in a displacement of the equilibrium position of $20\,\mathrm{fm}$.% Both can be neglected for our experimental conditions.

\begin{table}
	\caption{Important physical quantities used in this paper. \label{tab:parameters}}
	\begin{ruledtabular}
		\begin{tabular}{
				llp{58mm}
			}
			\textbf{Symbol}                                         & \textbf{Unit}    & \textbf{Description}                                                                                     \\ \hline
			$T_\mathrm{com}$                                        & $\mathrm{K}$     & Temperature of the \com{} motion                                                                         \\
			$T_\mathrm{bath}$                                       & $\mathrm{K}$     & Effective bath temperature of gas in vacuum chamber                                                      \\
			$T_\mathrm{int}$                                        & $\mathrm{K}$     & Internal temperature of the particle (in absence of $\mathrm{CO_2}$ laser illumination)                  \\
			$\Omega_0$                                              & $\mathrm{Hz}$    & Natural oscillation frequency                                                                            \\
			$\gamma$                                                & $\mathrm{Hz}$    & \com{} gas damping rate (inverse of relaxation time)                                                     \\
			$\Gamma$                                                & $\mathrm{Hz}$    & \com{} single-phonon heating rate due to interactions with the gas                                       \\
			$\Gamma^\prime$                                         & $\mathrm{K/s}$   & Gas energy heating rate (at which the \com{} temperature increases)                                      \\
			$I_\mathrm{CO_2}$                                       & $\mathrm{W/m^2}$ & Peak intensity of the $\mathrm{CO_2}$ laser in the focus (at the particle position)                      \\
			$\Delta\Omega_0$                                        & $\mathrm{Hz}$    & Shift of natural oscillation frequency due to increasing $T_\mathrm{int}$ with the $\mathrm{CO_2}$ laser \\
			$\Delta T_\mathrm{int}$                                 & $\mathrm{K}$     & Change of internal particle temperature due to $\mathrm{CO_2}$ laser irradiation                         \\
			$\alpha_\mathrm{G}$                                     & $1$              & Energy accommodation coefficient (energy transfer between gas and particle)                              \\
			$T_\mathrm{im}$ ($T_\mathrm{em}$)                         & $\mathrm{K}$     & Temperature of the gas molecules imping\-ing on (emerging from) the particle                           \\
			$\gamma_\mathrm{im}$ ($\gamma_\mathrm{em}$)               & $\mathrm{Hz}$    & \com{} damping rate caused by impinging (emerging) gas molecules                                       \\
			$\Gamma^\prime_\mathrm{im}$ ($\Gamma^\prime_\mathrm{em}$) & $\mathrm{K/s}$   & \com{} energy heating rate caused by impinging (emerging) gas molecules
		\end{tabular}
	\end{ruledtabular}
\end{table}

\section{Measurement of particle dynamics}

Using the optical levitation setup described above, we investigate the coupling between the particle's internal temperature and its \com{} motion. To measure the \com{} dynamics of the particle, we perform relaxation experiments that are described in Sec.~\ref{sec:relaxation-experiment}. In Sec.~\ref{sec:internal-temperature-change}, we determine the change of the particle's internal temperature $\Delta T_\mathrm{int}$ when activating the $\mathrm{CO_2}$ laser from the change of the particle's oscillation frequency $\Delta\Omega_0$. Both measurements, performed at varying gas pressures and at different $\mathrm{CO_2}$ laser intensities, allow us to deduce the absolute internal particle temperature $T_\mathrm{int}$ (see Sec.~\ref{sec:internal-particle-temperature}).

%\subsection{Relaxation experiment}\label{sec:relaxation-experiment}
\subsection{Measurement of center-of-mass temperature}\label{sec:relaxation-experiment}

\begin{figure}
	\includegraphics{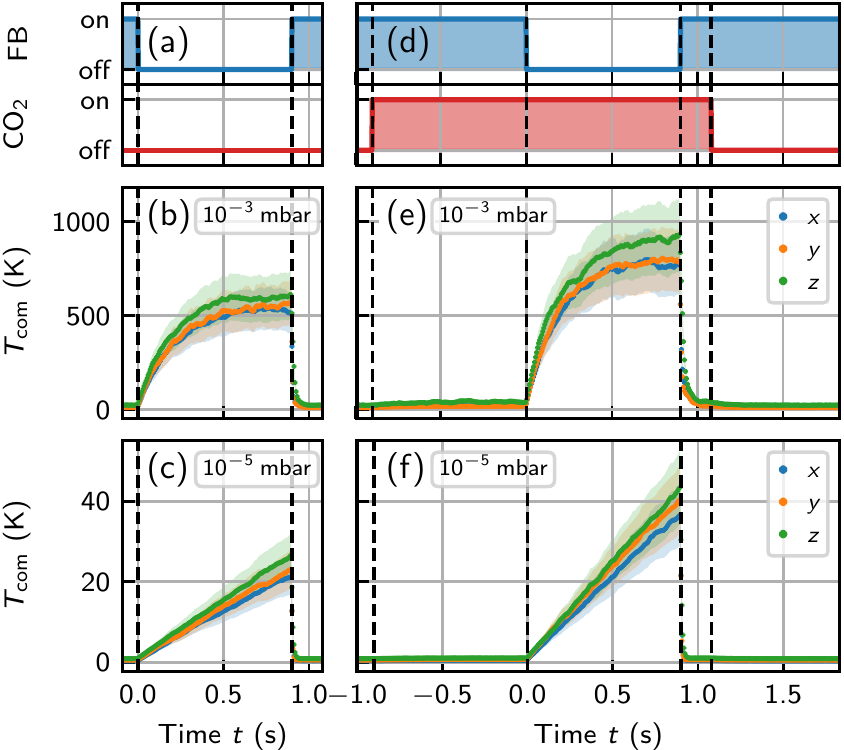}%
	\caption{Relaxation experiments. (a)~Using parametric feedback (FB) we first cool the particle's \com{} motion. We switch the feedback off at time $t=0$, such that the particle's average \com{} energy equilibrates towards the effective bath temperature. After $900\,\mathrm{ms}$, the feedback is turned on again to return to the initial conditions. (b)~At a pressure of $1\times10^{-3}\,\mathrm{mbar}$, we observe a relaxation of the \com{} temperature $T_\mathrm{com}$ to the bath temperature $T_\mathrm{bath}$ along all three oscillation axes. (c)~When reducing the pressure to $1\times10^{-5}\,\mathrm{mbar}$, we only observe the initial slope of the relaxation curve that rises linearly with the heating rate $\Gamma^\prime$. (d)~For measuring the particle dynamics at elevated internal temperatures, we activate the $\mathrm{CO_2}$ laser $900\,\mathrm{ms}$ before starting the relaxation experiment. (e)~At $10^{-3}\,\mathrm{mbar}$, in the presence of the $\mathrm{CO_2}$ heating laser, the particle's \com{} temperature equilibrates to a bath temperature higher than in the absence of the heating laser. (f)~At $10^{-5}\,\mathrm{mbar}$, the slope of the heating curve $\Gamma'$ is larger in the presence of the heating laser as compared to (c), where the heating laser is off. Quantities extracted from these measurements are summarized in Tab.~\ref{tab:reheating_params}. \label{fig:relaxation}}
\end{figure}

We investigate the coupling between the internal temperature and the \com{} motion of the particle by performing relaxation experiments. Each relaxation cycle starts with the particle under parametric feedback cooling along all three oscillation axes. We switch off the feedback and let the particle's \com{} motion relax from a feedback-cooled state with $T_\mathrm{com}\ll T_\mathrm{bath}$ to the bath temperature $T_\mathrm{bath}$ of the surrounding gas. From the evolution of the \com{} temperature, we extract the damping rate $\gamma$ and $T_\mathrm{bath}$, which describe the particle's dynamics according \rev{to} Eq.~(\ref{eq:eq-of-motion}). These two quantities determine the single-phonon heating rate $\Gamma=\gamma k_\mathrm{B}T_\mathrm{bath}/(\hbar\Omega_0)$, which is the rate at which phonons enter the \com{} oscillation mode. $\Gamma$ is a measure for the interaction strength of the particle's \com{} motion with its environment \cite{Gieseler2014a}.

In Fig.~\ref{fig:relaxation}(a), we show our experimental protocol for a single relaxation cycle in absence of $\mathrm{CO_2}$ laser heating, consisting of three steps: (1) initializing the particle's \com{} temperature $T_\mathrm{com}(t=0)$ well below the bath temperature $T_\mathrm{bath}$ with the parametric feedback (FB), (2) switching off the feedback at time $t=0$ to allow for a relaxation of the particle's \com{} temperature to the bath temperature, (3) reactivating the feedback after $900\,\mathrm{ms}$. During each relaxation cycle we record a time trace of the particle's position, from which we calculate \rev{its} kinetic energy in sections of $6\,\mathrm{ms}$ \cite{Hebestreit2017}. In Fig.~\ref{fig:relaxation}(b) we plot, for a pressure of $p_\mathrm{gas}=1\times10^{-3}\,\mathrm{mbar}$, the \com{} temperature that we obtain as an average over the kinetic energy of 1000 realizations of the relaxation cycle. We observe that after turning the parametric feedback off at $t=0$, the \com{} temperature in all three axes rises at \rev{the same} rate and asymptotically approaches a steady-state temperature of $T_\mathrm{bath}=571(21)\,\mathrm{K}$. Importantly, this bath temperature is significantly higher than the surrounding temperature of $300\,\mathrm{K}$. Thus, the bath temperature felt by the particle is not the surrounding temperature, but an \emph{effective} bath temperature.
We reason in Sec.~\ref{sec:two-bath-model-of-gas-interactions} that the increased bath temperature is caused by an elevated internal temperature of the particle. When switching the feedback on again, the \com{} temperature is cooled to $T_\mathrm{com}<10\,\mathrm{K}$, such that the system is initialized for the next experimental cycle.

The temporal evolution of the \com{} temperature is described by the rate equation \cite{Gieseler2014a,Jain2016}
\begin{equation}
\label{eq:temp_diffeq}
\frac{\mathrm{d}}{\mathrm{d}t}T_\mathrm{com}(t)=-\gamma T_\mathrm{com}(t)+\Gamma'.
\end{equation}
Here, for convenience, instead of using the single-phonon heating rate $\Gamma$ in units of $\mathrm{Hz}$, we define an energy heating rate $\Gamma^\prime=\Gamma\hbar\Omega_0/k_\mathrm{B}=\gamma T_\mathrm{bath}$ in units of $\mathrm{K/s}$. The first term on the right-hand side of Eq.~(\ref{eq:temp_diffeq}) describes the dissipation of energy from the \com{} oscillation, while the second term characterizes the rate at which energy enters the oscillation mode from the environment. For $\mathrm{d}T_\mathrm{com}/\mathrm{d}t=0$, we find that the \com{} temperature equals the bath temperature $T_\mathrm{bath}=\Gamma'/\gamma$. Solving Eq.~(\ref{eq:temp_diffeq}) yields the evolution from the initial \com{} temperature $T_\mathrm{com}(t=0)$ as
\begin{equation}
\label{eq:temp_relax}
T_\mathrm{com}(t)=T_\mathrm{bath}+\left[T_\mathrm{com}(0)-T_\mathrm{bath}\right]\mathrm{e}^{-\gamma t}.
\end{equation}
We fit Eq.~(\ref{eq:temp_relax}) to the measured relaxation curves shown in Fig.~\ref{fig:relaxation}(b) to extract the damping rate $\gamma/2\pi= 0.79(2)\,\mathrm{Hz}$, and derive a heating rate of $\Gamma'/(2\pi)=T_\mathrm{bath}\gamma/(2\pi)=451(29)\,\mathrm{K/s}$. We summarize all the measurement results from Fig.~\ref{fig:relaxation} in Tab.~\ref{tab:reheating_params}.

\begin{table}
	\caption{Experimental and fitting parameters for the measurements in Fig.~\ref{fig:relaxation}. The change in internal particle temperature $\Delta T_\mathrm{int}$ is determined according to Sec.~\ref{sec:internal-temperature-change} \label{tab:reheating_params}}
	\begin{ruledtabular}
		\begin{tabular}{
				ccccccc
			}
			Fig. & $p_\mathrm{gas}$  &     $I_\mathrm{CO_2}$      & $\Delta T_\mathrm{int}$ & $T_\mathrm{bath}$ & $\gamma/(2\pi)$ & $\Gamma^\prime/(2\pi)$ \\
			& $(\mathrm{mbar})$ & $\left(\mathrm{\frac{\mu W}{\mu m^2}}\right)$ &     $(\mathrm{K})$      &       $(\mathrm{K})$       & $(\mathrm{Hz})$ &    $\left(\mathrm{\frac{K}{s}}\right)$    \\ \hline
			2(b) & $1\times10^{-3}$         &            $0$             &           --            &          $571(21)$          &    $0.79(2)$    &       $451(29)$        \\
			2(e) & $1\times10^{-3}$ &           $0.47$            &        $601(16)$        &          $824(34)$          &    $0.88(2)$    &       $723(43)$        \\ \hline
			2(c) & $1\times10^{-5}$         &            $0$             &           --            &             --              &       --        &        $4.4(2)$        \\
			2(f) & $1\times10^{-5}$ &           $0.47$            &        $825(17)$        &             --              &       --        &        $7.3(3)$
		\end{tabular}
	\end{ruledtabular}
\end{table}

When lowering the gas pressure to $1\times10^{-5}\,\mathrm{mbar}$, as shown in Fig.~\ref{fig:relaxation}(c), we observe relaxation dynamics that are 100 times slower than those at $1\times10^{-3}\,\mathrm{mbar}$ such that letting the particle reheat close to its steady-state \com{} temperature would require $100\,\mathrm{s}$ for a single relaxation cycle and sufficiently many experimental cycles to average out the thermal fluctuations would take prohibitively long. Therefore, at low pressures, where $\gamma^{-1}\gg 1\,\mathrm{s}$, we only capture the linear beginning of the relaxation curve, where the \com{} temperature evolves as $T_\mathrm{com}(t)\approx T_\mathrm{com}(0)+\Gamma' t$. We find that the slope of the \com{} temperature in Fig.~\ref{fig:relaxation}(c) corresponds to a heating rate of $\Gamma'/(2\pi)=4.4(2)\,\mathrm{K/s}$. Comparing the heating rates obtained from the experiments shown in Figs.~\ref{fig:relaxation}(b) and (c) shows that $\Gamma'$ scales linearly with pressure. \rev{Given the linear pressure dependence of the damping rate $\gamma$, this suggests a pressure-independent bath temperature $T_\mathrm{bath}$, which can be attributed to the absence of convective cooling at pressures below $10^{-3}\,\mathrm{mbar}$ (see Sec.~\ref{sec:thermodynamic-estimations}) \cite{Gieseler2012,Jain2016}.}

Up to this point, we considered the particle in the absence of additional heating by $\mathrm{CO_2}$ laser irradiation. In order to study the influence of the internal particle temperature $T_\mathrm{int}$ on the dynamics, we repeat the relaxation experiments with an elevated internal particle temperature. Each relaxation cycle follows the sequence shown in Fig.~\ref{fig:relaxation}(d): We start \rev{with the} particle in a feedback-cooled state and activate the $\mathrm{CO_2}$ laser at $t=-0.9\,\mathrm{s}$. Experimentally, we find that the particle's internal temperature equilibrates within a few $100\,\mathrm{ms}$ (cf.~Sec.~\ref{sec:internal-temperature-change}) \cite{Bateman2014}. We therefore wait for $900\,\mathrm{ms}$, before switching off the feedback at time $t=0$ for $900\,\mathrm{ms}$. After the relaxation cycle, we switch off the $\mathrm{CO_2}$ laser at time $t=1.08\,\mathrm{s}$ to let the particle's internal temperature cool down to its initial value. In Fig.~\ref{fig:relaxation}(e), we show the temperature relaxation along the three axes of the \com{} motion at a pressure of $1\times10^{-3}\,\mathrm{mbar}$ for a particle heated with the $\mathrm{CO_2}$ laser. We observe a relaxation to a \com{} temperature of $T_\mathrm{bath}=824(34)\,\mathrm{K}$, which is higher than that observed in the absence of additional heating [cf.~Fig.~\ref{fig:relaxation}(b)]. This shows that increasing the internal particle temperature effectively causes a rise in the extracted bath temperature $T_\mathrm{bath}$. Furthermore, we measure a higher damping rate $\gamma$ in the case with additional heating, which results in an increase of the heating rate $\Gamma^\prime$ by $60(14)\%$ compared to Fig.~\ref{fig:relaxation}(b) in the absence of $\mathrm{CO_2}$ laser heating (see Tab.~\ref{tab:reheating_params}). Accordingly, an increased internal temperature $T_\mathrm{int}$ of the trapped particle gives rise to an increased (effective) bath temperature $T_\mathrm{bath}$, entailing an increased heating rate $\Gamma'$.

In Fig.~\ref{fig:relaxation}(f), we plot the measured relaxation curves for the internally heated particle at $1\times 10^{-5}\,\mathrm{mbar}$. We observe, in agreement with the results at $1\times10^{-3}\,\mathrm{mbar}$, that the heating rate $\Gamma^\prime$ increases by $66(10)\%$ in comparison with the results in Fig.~\ref{fig:relaxation}(c) for the non-heated particle (see Tab.~\ref{tab:reheating_params}).
%Note that the time axis in Fig.~\ref{fig:relaxation}(b,d) corresponds to the one in Fig.~\ref{fig:temperature}(b). Accordingly, the measurement of Fig.~\ref{fig:temperature}(b) confirms that an equilibrium of the internal temperature is reached before the relaxation experiment is started.

%\subsection{Heating Rates}\label{sec:heating-rates}

\begin{figure}
	\includegraphics{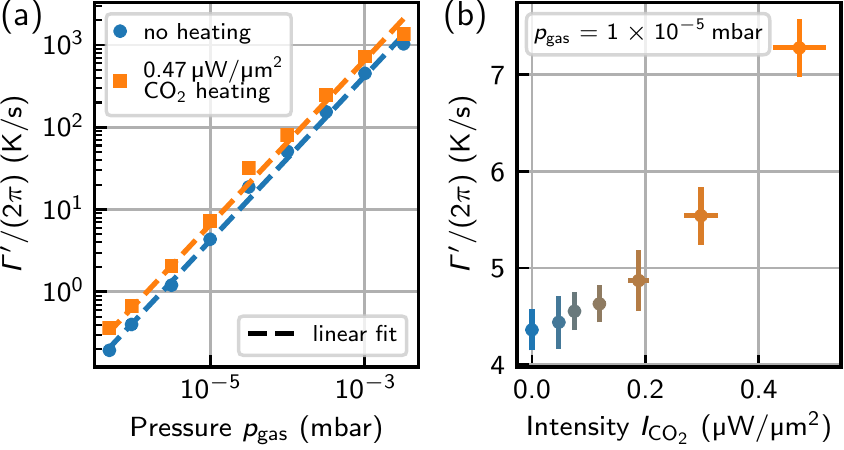}%
	\caption{(a)~Heating rates as a function of pressure extracted from relaxation measurements without additional heating of the internal temperature (blue) and with maximal heating ($\mathrm{CO_2}$ laser intensity of $I_\mathrm{CO_2}=0.47\,\mathrm{\mu W/\mu m^2}$, orange). The dashed lines are linear fits to the data points. Error bars are smaller than the marker size. (b)~Heating rates at different intensities of the $\mathrm{CO_2}$ laser measured at a pressure of $1\times10^{-5}\,\mathrm{mbar}$. The error bars indicate the standard deviation of the measurements. \label{fig:heating_rate}}
\end{figure}

To determine the particle's \com{} dynamics at different pressures and internal temperatures, we perform relaxation measurements in a pressure range between $5\times 10^{-7}$ and $3\times 10^{-3}\,\mathrm{mbar}$. 
%At higher pressures the parametric feedback is not able to efficiently reduce the particle's \com{} temperature, making relaxation measurements impossible. 
Furthermore, we vary the intensity in the focus of the $\mathrm{CO_2}$ laser from $0$ to $0.47\,\mathrm{\mu W/\mu m^2}$. \rev{To account for possible drifts of the experimental setup, we re-calibrated our detectors at $1\times 10^{-3}$ and $1\times 10^{-5}\,\mathrm{mbar}$.} In Fig.~\ref{fig:heating_rate}(a), we summarize the measured heating rates for different pressures. Below $10^{-3}\,\mathrm{mbar}$, the heating rate depends linearly on the gas pressure. Furthermore, the heating rates under illumination with the $\mathrm{CO_2}$ laser (orange) are consistently higher than for the case without increasing the internal temperature (blue). The dependence of the heating rate on the peak intensity of the $\mathrm{CO_2}$ laser is shown in Fig.~\ref{fig:heating_rate}(b). Our measurements show a clear dependence of the \com{} dynamics on  $\mathrm{CO_2}$ laser irradiation, which suggests a coupling between the \com{} motion and the particle's internal temperature. %To quantify this coupling and to verify the heating of the internal particle temperature by the $\mathrm{CO_2}$ laser, in the next section, we measure the change in the internal temperature.

%By measuring the heating rate at varying pressures, we found a clear dependence of the \com{} dynamics on the $\mathrm{CO_2}$ laser irradiation, which indicates a coupling between the \com{} motion and the particle's internal temperature. To quantify this coupling, we require information on how much the $\mathrm{CO_2}$ laser increases the internal particle temperature. We measure this change in the internal temperature in the following Sec.~\ref{sec:internal-temperature-change} using the temperature dependent frequency shift of the particle's oscillation frequency.

%\subsection{Internal temperature change}\label{sec:internal-temperature-change}
\subsection{Measurement of internal temperature change}\label{sec:internal-temperature-change}

\begin{figure}
	\includegraphics{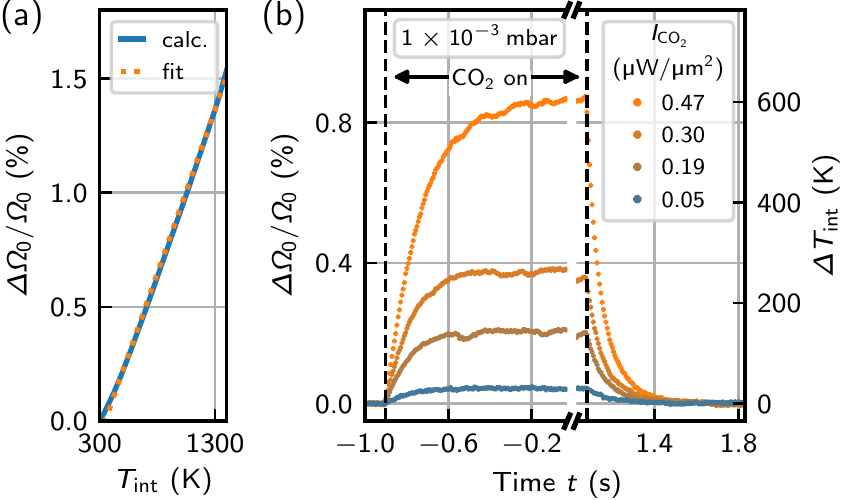}%
	\caption{(a)~Calculated temperature dependence of oscillation frequency. When the particle is heated with the $\mathrm{CO_2}$ laser, its oscillation frequency increases due to changes in the particle's material properties. This leads to a nearly linear relation between relative frequency change and increase of the internal particle temperature with $\Delta \Omega_0/\Omega_0=(1.43\times10^{-5}\,\mathrm{K}^{-1})\Delta T_\mathrm{int}$\rev{, where $\Omega_0$ denotes the oscillation frequency at an internal particle temperature of $300\,\mathrm{K}$}. (b)~Measured relative frequency shift due to heating of particle with $\mathrm{CO_2}$ laser for different heating laser powers at a pressure of $1\times10^{-3}\,\mathrm{mbar}$. \rev{Here, $\Omega_0$ is the particle oscillation frequency without $\mathrm{CO_2}$ laser irradiation.} The time axis corresponds to the one shown in the relaxation cycle in Fig.\ \ref{fig:relaxation}(e). According to the functional dependence extracted from (a), the maximally observed frequency shift corresponds to a temperature rise of $\Delta T_\mathrm{int}=601(16)\,\mathrm{K}$. \label{fig:temperature}}
\end{figure}

Increasing the internal particle temperature in a controlled way with the $\mathrm{CO_2}$ laser enables us to study the dependence of the particle dynamics on the internal particle temperature $T_\mathrm{int}$. For a quantitative analysis, we additionally require an independent measurement of the increase of the internal particle temperature $\Delta T_\mathrm{int}$ due to the $\mathrm{CO_2}$ laser irradiation. 
%A direct absolute measurement of the internal temperature is challenging. However, we are able to 
We deduce the change in the internal temperature using the temperature dependence of the particle's material properties. In particular, we exploit the temperature dependence of the \rev{mass} density $\rho$ \cite{Brueckner1970} and the refractive index $n$ \cite{Waxler1973} of silica. These material properties lead to a temperature dependent oscillation frequency $\Omega_0$ of the particle, which scales as $\Omega_0\propto\sqrt{\rho^{-1}(n^2-1)/(n^2+2)}$ \cite{Gieseler2012}. In Fig.~\ref{fig:temperature}(a), we plot as a blue line the calculated relative shift of the particle's oscillation frequency $\Delta \Omega_0/\Omega_0$ as a function of the internal particle temperature $T_\mathrm{int}$ according to the material data listed in Refs.~\cite{Waxler1973, Brueckner1970}, where $\Omega_0$ denotes the oscillation frequency at an internal temperature of $300\,\mathrm{K}$. The relative oscillation-frequency shift depends almost linearly on the internal particle temperature $T_\mathrm{int}$ and we find a proportionality factor $\Delta\Omega_{0}/\Omega_{0} = 1.43(1)\times 10^{-5}\,\mathrm{K^{-1}}\times\Delta T_\mathrm{int}$. Accordingly, we can determine the relative increase in internal temperature of the particle $\Delta T_\mathrm{int}$ caused by the irradiation by the $\mathrm{CO_2}$ laser from the relative shift of the particle oscillation frequency. This assumes that the internal temperature is homogeneous across the particle \cite{Millen2014}. Furthermore, since there is no data available for the refractive index $n$ of glass above $1000\,\mathrm{K}$, we extend the linear trend of the relative frequency change found in Fig.~\ref{fig:temperature}(a) to a temperature of $1800\,\mathrm{K}$, the highest internal particle temperature occurring in this work.

In Fig.~\ref{fig:temperature}(b), we show the measured relative change in oscillation frequency as a function of time when heating the particle with different $\mathrm{CO_2}$ laser intensities at a pressure of $1\times10^{-3}\,\mathrm{mbar}$. We perform repeated heating cycles, where we start with the particle's \com{} motion being feedback cooled and the heating laser being switched off. We activate the $\mathrm{CO_2}$ laser from $t=-0.9\,\mathrm{s}$ to $1.08\,\mathrm{s}$ [cf.~Fig.~\ref{fig:relaxation}(d,e)], in order to illuminate the particle with a $\mathrm{CO_2}$ laser intensity of up to $I_\mathrm{CO_2}=0.47\,\mathrm{\mu W/\mu m^2}$. During the entire heating cycle, we record the \com{} position of the particle along the three axes, extract the oscillation frequency from segments of $6\,\mathrm{ms}$ length, and average each segment over 1000 realizations. We calculate the relative change of the oscillation frequency and average over the three oscillation axes. After activating the $\mathrm{CO_2}$ laser, we find an up-shift of the oscillation frequency of up to $0.9\%$ for the maximum $\mathrm{CO_2}$ laser intensity. We interpret this frequency shift as an increase of the internal particle temperature due to absorption heating. Applying the linear conversion factor of $1.43\times 10^{-5}\,\mathrm{K^{-1}}$ found in Fig.~\ref{fig:temperature}(a), we obtain a change in the particle's internal temperature of $\Delta T_\mathrm{int}=601(16)\mathrm{K}$. Using higher $\mathrm{CO_2}$ laser intensities causes particle loss during the heating cycle. We attribute these losses to strong changes in material properties when reaching the working point of glass, or to the high \com{} oscillation amplitudes due to the elevated internal temperature of the particle. %\footnote{\color{red} This can easily be checked in the lab by illuminating the particle with high laser powers at low pressures (lower amplitudes throughout the experiment) and without turning off the feedback.}
%Using this method we reach exceptionally high resolution of $30\,\mathrm{ppm}$ for the relative change in oscillation frequencies in $x$ and $y$ direction, which corresponds to a precision in detecting internal particle temperature changes of less than $3\,\mathrm{K}$.

\section{Two-bath model of gas interactions}\label{sec:two-bath-model-of-gas-interactions}

Having established experimental methods to measure the particle's \com{} dynamics and the change of its internal temperature $\Delta T_\mathrm{int}$ due to $\mathrm{CO_2}$ laser irradiation, in the following, we 
develop a theoretical model 
%use the measured heating rates 
to extract the 
%unknown absolute 
internal particle temperature $T_\mathrm{int}$ at any gas pressure.
% below $10^{-3}\,\mathrm{mbar}$. 
We describe the interaction of an internally hot particle with the surrounding gas following the model developed in Ref.~\cite{Millen2014}. It assumes an {\em inelastic} scattering process, that is,
an impinging gas molecule with temperature $T_\mathrm{im}$  heats up to a temperature $T_\mathrm{em}$ upon interacting with the hot particle. As a consequence, the particle interacts with two different baths, the bath of impinging molecules with parameters $T_\mathrm{im}$, $\gamma_\mathrm{im}$, $\Gamma'_\mathrm{im}$, and the bath of emerging gas molecules with parameters $T_\mathrm{em}$, $\gamma_\mathrm{em}$, $\Gamma'_\mathrm{em}$ \cite{Epstein1924}. Therefore, the total gas damping rate is $\gamma = \gamma_\mathrm{im}+\gamma_\mathrm{em}$ and the total gas heating rate equals $\Gamma'=\Gamma'_\mathrm{im}+\Gamma'_\mathrm{em}=T_\mathrm{im}\gamma_\mathrm{im}+T_\mathrm{em}\gamma_\mathrm{em}$ \cite{Millen2014}. The damping rate associated with the emerging gas molecules is $\gamma_\mathrm{em}=\pi\sqrt{T_\mathrm{em}/T_\mathrm{im}}\gamma_\mathrm{im}/8$ and the damping rate due to the impinging gas molecules is $\gamma_\mathrm{im}=32R^2p_\mathrm{gas}/(3m\bar{v}_\mathrm{gas})$. Here,  $\bar{v}_\mathrm{gas}=\sqrt{8N_\mathrm{A}k_\mathrm{B}T_\mathrm{im}/(\pi M)}$ is the mean velocity of gas molecules, where $M$ is the molar mass of the gas and $N_\mathrm{A}$ is Avogadro's constant \cite{Millen2014}. \rev{Introducing} the energy accommodation coefficient $\alpha_G=(T_\mathrm{em}-T_\mathrm{im})/(T_\mathrm{int}-T_\mathrm{im})$, which describes the energy transfer from the particle to the gas, we find that the total  heating rate follows
%, which we find to be $\alpha_G=0.65$ at our experimental conditions (see Sec.~\ref{sec:int-temp-and-accommodation-coeff}). 
\begin{equation}\label{eq:heating_rate}
\Gamma'=\gamma_\mathrm{im}\cdot\left\{T_\mathrm{im}+\frac{\pi}{8}\frac{\left[\alpha_G T_\mathrm{int} + (1-\alpha_G)T_\mathrm{im}\right]^{3/2}}{\sqrt{T_\mathrm{im}}}\right\}\; .
\end{equation}
It depends on the particle's properties (radius, mass, and internal temperature), the gas properties (pressure, temperature, and composition), and the energy transfer between particle and gas. This two-bath model predicts that increasing the particle's internal temperature $T_\mathrm{int}$ causes a rise of the gas heating rate $\Gamma'$, which is what we observe experimentally in Fig.~\ref{fig:relaxation}. In particular, as described in the following section, we are able to derive the internal temperature of the particle using the measured heating rates (c.f. Fig.~\ref{fig:heating_rate}a). %In the following section, we discuss this determination of the internal particle temperature in two different ways, which eventually allows us to extract both, the internal particle temperature and the energy accommodation coefficient from our measurement data.

%At high vacuum the gas molecules impinging on the particle thermalize to the environment temperature $T_\mathrm{im}=T_\mathrm{env}=300\,\mathrm{K}$. The emerging gas molecules will have the elevated temperature $T_\mathrm{em}=\alpha_G T_\mathrm{int}+(1-\alpha_G)T_\mathrm{im}$. Using this theory we can predict the gas heating rate that the particle experiences, indicated by the dashed lines in Fig.~\ref{fig:heating_rate}(b). Here we assumed an internal particle temperature of $T_\mathrm{int}=1600\,\mathrm{K}$ according to a theoretical model, that we will verify experimentally later. The dotted line in Fig.~\ref{fig:heating_rate}(b) illustrates the recoil heating rate according to \cite{Jain2016}. The total heating rate according to our models is shown as the solid lines.

\section{Internal particle temperature}\label{sec:internal-particle-temperature}

%Using the two-bath model of the particle's interaction with the gas, we are able to determine the absolute internal temperature of the particle $T_\mathrm{int}$ from the measured heating rates [Fig.~\ref{fig:heating_rate}(a)]. 
According to Eq.~(\ref{eq:heating_rate}), the heating rate depends on the internal temperature of the particle $T_\mathrm{int}$, the energy accommodation coefficient $\alpha_\mathrm{G}$, the molar mass of the gas $M$ and the temperature of the impinging gas molecules $T_\mathrm{im}$, which equals the temperature of the vacuum chamber walls and is measured throughout the experiment. To determine the molar mass $M$ of the gas, we measure the gas composition at pressures below $10^{-3}\,\mathrm{mbar}$ with a residual gas analyzer. We find that the gas consists to $58(9)\%$ of water vapor, and contains in addition nitrogen, hydrogen, oxygen and carbon dioxide. We deduce an average molar mass of $M=20(11)\,\mathrm{g/mol}$. The energy accommodation coefficient $\alpha_G$ also depends on the gas composition in the vacuum chamber. However, as the energy accommodation coefficient of water vapor on silica is not known, for now, we assume a value of $\alpha_G=0.65$. At the end of this section, we justify this choice by determining the accommodation coefficient $\alpha_G$ from the experimental data.

\begin{figure}
	\includegraphics{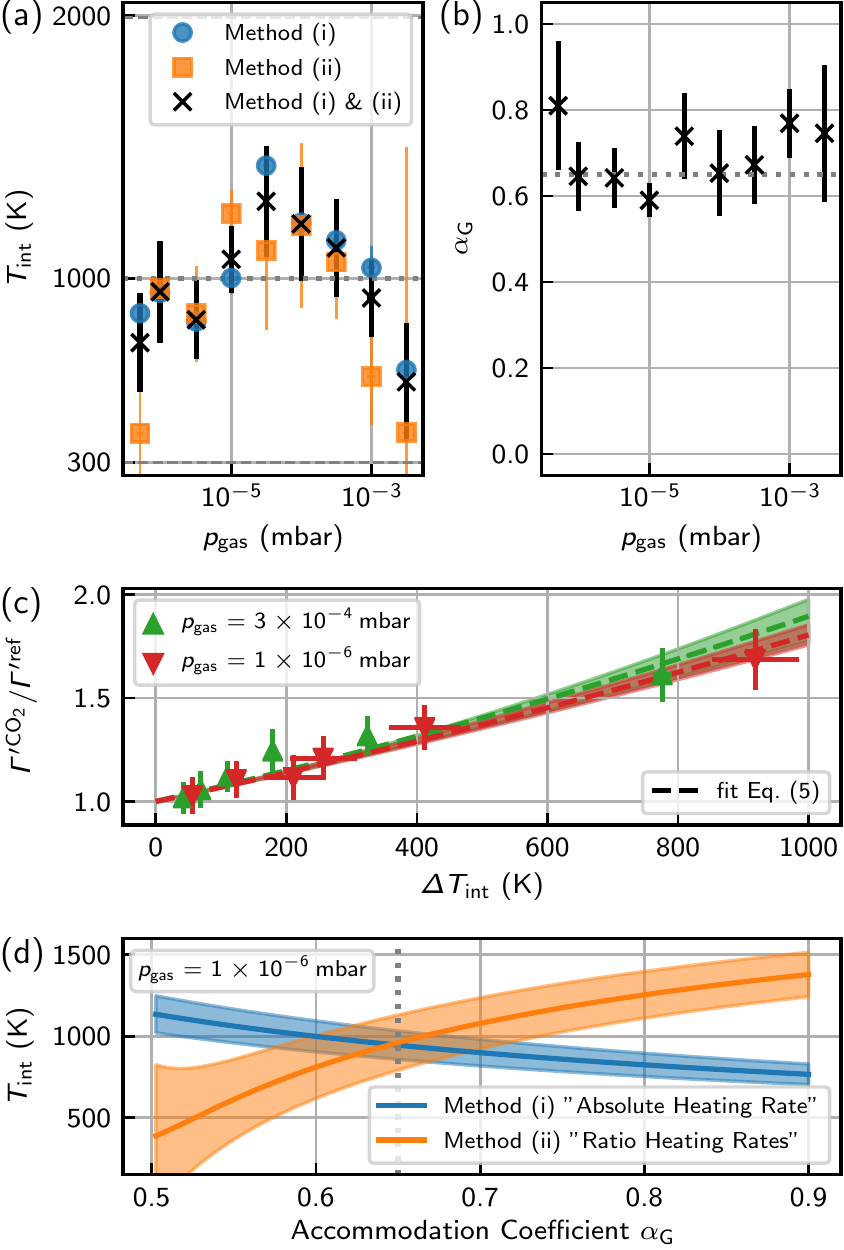}%
	\caption{(a)~Internal particle temperatures derived using 
	%three 
	different methods: \emph{(i)} using the absolute heating rate of the internally non-heated particle (blue), \emph{(ii)} using the ratio of heating rates for an internally heated and internally non-heated particle (orange). The combination of method \emph{(i)} and \emph{(ii)} is shown in black [see (d)]. Marked are room temperature ($300\,\mathrm{K}$), the melting temperature of silica ($\sim2000\,\mathrm{K}$), and the average of the internal temperatures $T_\mathrm{int}=1000(60)\,\mathrm{K}$ at pressures below $10^{-3}\,\mathrm{mbar}$.
	(b)~Corresponding to the combined method in (a), we plot the deduced energy accommodation coefficients and mark \rev{their} mean value $\alpha_\mathrm{G}=0.65(3)$.
	(c)~To illustrate method \emph{(ii)}, we plot the ratio of the heating rates with and without the $\mathrm{CO_2}$ laser illuminating the particle as a function of $\Delta T_\mathrm{int}$ for two pressures. To each dataset, we fit Eq.~(\ref{eq:ratio-heating-rates}) and infer an internal particle temperature $T_\mathrm{int}$ as a fit parameter.
	(d)~Combining method \emph{(i)} and \emph{(ii)}, we plot the derived internal particle temperature depending on the chosen accommodation coefficient for methods \emph{(i)} and \emph{(ii)} at a pressure of $1\times10^{-6}\,\mathrm{mbar}$. From the intersection point, we infer the best-fit energy accommodation coefficient and the internal particle temperature [see (a,b), black data points].
	The error bars and shaded areas in all plots indicate the $1\sigma$ confidence intervals.\label{fig:internal_temperature}}
\end{figure}

We deduce the internal particle temperature $T_\mathrm{int}$ from the measured heating rates in two ways: \emph{(i)} from the absolute measured heating rates $\Gamma^\prime$ in the absence of $\mathrm{CO_2}$ laser heating, and \emph{(ii)} using the heating rate measured at varying internal temperatures $T_\mathrm{int}+\Delta T_\mathrm{int}$, where the change in the internal temperature is known from the oscillation frequency shift (see Sec.~\ref{sec:internal-temperature-change}). For method \emph{(i)}, we solve Eq.~(\ref{eq:heating_rate}) for $T_\mathrm{int}$ and use the heating rates measured in Fig.~\ref{fig:heating_rate}(a) (blue data points). In Fig.~\ref{fig:internal_temperature}(a), we plot in blue the internal temperature that we extract with this method for different gas pressures. We obtain temperature values that are distributed between $600$ and $1400\,\mathrm{K}$.

For method \emph{(ii)}, we compute the ratio of the measured heating rate $\Gamma^{\prime\mathrm{CO_2}}$ in the presence of the $\mathrm{CO_2}$ heating laser and the heating rate $\Gamma^{\prime\mathrm{ref}}$ in absence of the heating laser. From Eq.~(\ref{eq:heating_rate}), we find the theoretical expression
\begin{equation}\label{eq:ratio-heating-rates}
\frac{\Gamma^{\prime\mathrm{CO_2}}}{\Gamma^{\prime\mathrm{ref}}} = \frac{\frac{8}{\pi} T_\mathrm{im}^{3/2}+\left(T_\mathrm{em}^\mathrm{CO_2}\right)^{3/2}}{\frac{8}{\pi} T_\mathrm{im}^{3/2}+\left(T_\mathrm{em}^\mathrm{ref}\right)^{3/2}}.
\end{equation}
The temperature of the emerging gas molecules in the case with the activated $\mathrm{CO_2}$ laser is $T_\mathrm{em}^\mathrm{CO_2}=\alpha_\mathrm{G}(T_\mathrm{int}+\Delta T_\mathrm{int})+(1-\alpha_\mathrm{G})T_\mathrm{im}$, and without additional heating $T_\mathrm{em}^\mathrm{ref}=\alpha_\mathrm{G} T_\mathrm{int}+(1-\alpha_\mathrm{G})T_\mathrm{im}$. We measure the change of internal temperature $\Delta T_\mathrm{int}$ using the shift in oscillation frequency as described in Sec.~\ref{sec:internal-temperature-change}. Accordingly, we can deduce the absolute internal particle temperature $T_\mathrm{int}$ as the only free parameter in Eq.~(\ref{eq:ratio-heating-rates}). % In particular, this method does not require a calibrated detector signal.
We plot the ratio of the heating rates with and without the $\mathrm{CO_2}$ laser in Fig.~\ref{fig:internal_temperature}(c) at $3\times10^{-4}\,\mathrm{mbar}$ (green) and at $1\times10^{-6}\,\mathrm{mbar}$ (red) for 6 different $\mathrm{CO_2}$ laser intensities. To these data points we fit Eq.~(\ref{eq:ratio-heating-rates}) (dashed lines). The shaded areas indicate the $1\sigma$ confidence interval of the fit. The only free parameter for the fits is the absolute internal temperature of the particle $T_\mathrm{int}$ in the absence of $\mathrm{CO_2}$ laser irradiation. We plot these internal particle temperatures that were extracted with method \emph{(ii)} in Fig.~\ref{fig:internal_temperature}(a) in orange and find values in a similar range like for method \emph{(i)}.% In contrast to the first method, using the ratios of the heating rates eliminates the need of a calibration of the detector signal, which may drift over time.

%\subsection{Measurement of Internal Temperature and Accommodation Coefficient}\label{sec:int-temp-and-accommodation-coeff}

So far, we have discussed two independent methods to determine the internal temperature of the particle using the absolute value of the heating rate, or the ratio of the heating rates at different internal temperatures. However, in both cases we had to assume a value for the \emph{a priori} unknown energy accommodation factor $\alpha_\mathrm{G}$. We now combine both methods to deduce both the internal particle temperature and the energy accommodation coefficient from the experimental data. In Fig.~\ref{fig:internal_temperature}(d), we show for a pressure of $1\times10^{-6}\,\mathrm{mbar}$ the calculated internal particle temperature determined using the two methods for varying accommodation coefficients between $\alpha_\mathrm{G}=0.5$ and $0.9$. While methods \emph{(i)} and \emph{(ii)} are independent, they should both yield the same result for the internal particle temperature. Accordingly, the intersection of both graphs in Fig.~\ref{fig:internal_temperature}(d) provides the accommodation coefficient, where the results of both methods coincide. In this example at $1\times10^{-6}\,\mathrm{mbar}$, we find $T_\mathrm{int}=950(200)\,\mathrm{K}$ without the $\mathrm{CO_2}$ laser activated, and $\alpha_\mathrm{G}=0.65(8)$. In Fig.~\ref{fig:internal_temperature}(a), we plot in black the internal particle temperature that we determine using the combined method at the different gas pressures. \rev{We compare these measurements with the corresponding model in Sec.~\ref{sec:thermodynamic-estimations}.} In addition, in Fig.~\ref{fig:internal_temperature}(b), we show the corresponding energy accommodation coefficient for the different gas pressures. For the energy accommodation coefficient, we do not find a significant trend when lowering the gas pressure, suggesting that in our setup below $10^{-3}\,\mathrm{mbar}$ the relative gas composition does not depend on pressure. This \rev{conclusion} is supported by our measurement of the gas composition using a residual gas analyzer. Accordingly, we deduce a mean energy accommodation factor of $\alpha_\mathrm{G}=0.65(3)$ by averaging over all the measurements in Fig.~\ref{fig:internal_temperature}(b).

\section{Thermodynamic Estimations}\label{sec:thermodynamic-estimations}

Now that we are able to measure the internal particle temperature 
%at pressures below $10^{-3}\,\mathrm{mbar}$, 
we compare our  results with theoretical thermodynamic estimations. We consider the following thermal absorption and emission processes that the particle experiences: The particle's internal energy increases due the to absorption of laser light and incident blackbody radiation from the environment. On the other hand, the particle dissipates internal energy through blackbody radiation and gas convection \cite{Chang2010}. In thermodynamic equilibrium, we derive a steady-state internal temperature of the particle $T_\mathrm{int}$.

\begin{figure}
	\includegraphics{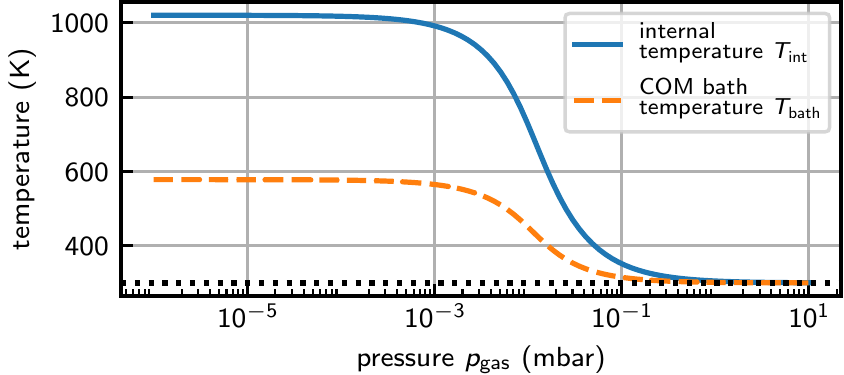}%
	\caption{Internal particle temperature calculated as a function of pressure from the balance of emission and absorption rates (blue solid line). At pressures above $10^{-1}\,\mathrm{mbar}$, convective gas cooling dominates the dissipation of internal energy and the internal particle temperature equilibrates to the environment temperature of $300\,\mathrm{K}$ (black dotted line). Below $10^{-3}\,\mathrm{mbar}$, convection is not efficient anymore and the internal temperature approaches $1020\,\mathrm{K}$. Additionally, we calculate the resulting \com{} equilibrium temperature $T_\mathrm{bath}$ from the two-bath model (Sec.~\ref{sec:two-bath-model-of-gas-interactions}) using the parameters determined in the experiments (orange dashed line). \label{fig:thermodynamics}}
\end{figure}

In Fig.~\ref{fig:thermodynamics}, we plot the internal temperature for varying gas pressure that we calculate following Ref.~\cite{Chang2010} (blue solid line). Here, we assume an accommodation coefficient of $\alpha_\mathrm{G}=0.65$ as we measured before in Sec.~\ref{sec:internal-particle-temperature}, and the refractive index of silica listed in Ref.~\cite{Kitamura2007}. This calculation shows that there is a transition from low internal temperatures at pressures above $1\,\mathrm{mbar}$, where the dissipation of internal energy is dominated by gas convection, to high internal temperatures below $10^{-3}\,\mathrm{mbar}$, where the only efficient dissipation mechanism is blackbody emission \cite{Ranjit2015a}. For our parameters, the equilibrium temperature at pressures below $10^{-3}\,\mathrm{mbar}$ stabilizes at $T_\mathrm{int}=1020\,\mathrm{K}$. Therefore, we also expect a constant internal particle temperature for gas pressures below $10^{-3}\,\mathrm{mbar}$ in our experiment. Computing the weighted average of the measured internal particle temperature below $10^{-3}\,\mathrm{mbar}$ from Fig.~\ref{fig:internal_temperature}(a) (black crosses), we find $T_\mathrm{int}=1000(60)\,\mathrm{K}$, which is in agreement with the thermodynamic estimate. \rev{We attribute the deviations of the data in Fig.~5(a) from this model to measurement uncertainty and drift of the calibration.}

Additionally, we plot in Fig.~\ref{fig:thermodynamics} the expected effective \com{} bath temperature $T_\mathrm{bath}=(T_\mathrm{im}\gamma_\mathrm{im}+T_\mathrm{em}\gamma_\mathrm{em})/(\gamma_\mathrm{im}+\gamma_\mathrm{em})$ (orange dashed line) for our experimental parameters and the energy accommodation factor of $\alpha_G=0.65$ that we previously deduced from our measurements. The bath temperature approaches a value of $580\,\mathrm{K}$ at high vacuum, where convective cooling through the residual gas becomes negligible. This theoretical result is in agreement with our measurement in Fig.~\ref{fig:relaxation}(b).

%In Fig.~\ref{fig:thermodynamics}~(b,c), we check the dependency of the thermodynamic calculations on two physical parameters. First, we vary the accommodation coefficient $\alpha_\mathrm{G}$ and find that it does not influence the equilibrium internal temperature at low pressures, but leads to a shift of the pressure in which we observe the temperature transition. A low accommodation coefficient causes a less efficient convective cooling of the particle, such that the particle heats up more at higher pressures as compared to a higher accommodation coefficient. Another parameter that we do not know very well is the effective blackbody permittivity. In Fig.~\ref{fig:thermodynamics}~(c), we vary the imaginary part of $n_\mathrm{b}$ (while leaving the refractive index at the trapping wavelength constant) and find that it greatly influences the equilibrium temperature at low pressures. This can be explained, because at low pressures (less than $10^{-3}\,\mathrm{mbar}$) the heat dissipation is dominated by blackbody emission. A larger imaginary part of the refractive index $n_b$ enhances the blackbody emission at low pressures and leaves the particle at a colder internal temperature.

\section{Conclusion}
We have shown that the coupling between the internal degrees of freedom of a levitated nanoparticle and its \com{} motion persists in a gas pressure range from $10^{-3}\,\mathrm{mbar}$ to $10^{-6}\,\mathrm{mbar}$. This coupling causes an elevated effective bath temperature above room temperature at high vacuum, which is of importance for levitated particle sensors as it decreases the sensitivity compared to an equivalent sensor at room temperature \cite{Ranjit2015a,Ranjit2016}. We expect this coupling to be present also for lower pressures, up to a point where interactions with the photon bath start to dominate the \com{} dynamics of the levitated particle \cite{Jain2016}. In this lower pressure regime, other coupling mechanisms between the internal and \com{} temperature will become important, e.g., mediated through the recoil of blackbody photons that depend on the internal particle temperature.

Furthermore, using the gas-mediated coupling and applying a model of the interaction of the particle with the surrounding dilute gas, we estimate the internal particle temperature. For our system, we deduce a mean internal particle temperature of $T_\mathrm{int}=1000(60)\,\mathrm{K}$ at gas pressures below $10^{-3}\,\mathrm{mbar}$ and an energy accommodation coefficient of $\alpha_\mathrm{G}=0.65(3)$. This high internal particle temperature is mainly caused by absorptive heating of the trapping laser and weak emission of blackbody radiation. We point out that our method is not limited to determining the internal temperature of optically levitated silica nanoparticles, but is universally applicable to any other material system that can be levitated and \com{} cooled.

In view of future quantum interference experiments, which require internal particle temperatures well below room temperature to achieve superpositions larger than the particle diameter, our measurements underline the importance of developing methods to control the internal particle temperature, as decoherence due to blackbody radiation prevents the appearance of interference patterns for high internal particle temperatures \cite{Romero-Isart2011,Kaltenbaek2012,Arndt2014}. Cooling of the internal particle temperature, as demonstrated at high gas pressure for levitated systems with an internal level structure, could serve as a key technology to reduce those decoherence effects for optically levitated particles \cite{Rahman2017}. \rev{As the internal particle temperature is determined by the absorption of the trapping laser, one may consider combining different particle materials with suitable trapping wavelengths, such as silicon and telecom wavelength \cite{Bateman2014}.} Further possibilities to reduce the internal particle temperature include alternative levitation strategies, such as Paul traps \cite{Millen2015} and magnetic levitation \cite{Romero-Isart2012}.

\begin{acknowledgments}
The authors acknowledge valuable discussions with C.~Dellago, O.~Romero-Isart, and J.~Gieseler. We thank P.~Kurpiers and A.~Wallraff for lending us their residual gas analyzer. This work has been supported by ERC-QMES (No.~338763) and the Swiss National Centre of Competence in Research (NCCR) - Quantum Science and Technology (QSIT) program (No.~51NF40-160591).
\end{acknowledgments}

%\bibliography{heating_references}

\begin{thebibliography}{32}%
\makeatletter
\providecommand \@ifxundefined [1]{%
 \@ifx{#1\undefined}
}%
\providecommand \@ifnum [1]{%
 \ifnum #1\expandafter \@firstoftwo
 \else \expandafter \@secondoftwo
 \fi
}%
\providecommand \@ifx [1]{%
 \ifx #1\expandafter \@firstoftwo
 \else \expandafter \@secondoftwo
 \fi
}%
\providecommand \natexlab [1]{#1}%
\providecommand \enquote  [1]{``#1''}%
\providecommand \bibnamefont  [1]{#1}%
\providecommand \bibfnamefont [1]{#1}%
\providecommand \citenamefont [1]{#1}%
\providecommand \href@noop [0]{\@secondoftwo}%
\providecommand \href [0]{\begingroup \@sanitize@url \@href}%
\providecommand \@href[1]{\@@startlink{#1}\@@href}%
\providecommand \@@href[1]{\endgroup#1\@@endlink}%
\providecommand \@sanitize@url [0]{\catcode `\\12\catcode `\$12\catcode
  `\&12\catcode `\#12\catcode `\^12\catcode `\_12\catcode `\%12\relax}%
\providecommand \@@startlink[1]{}%
\providecommand \@@endlink[0]{}%
\providecommand \url  [0]{\begingroup\@sanitize@url \@url }%
\providecommand \@url [1]{\endgroup\@href {#1}{\urlprefix }}%
\providecommand \urlprefix  [0]{URL }%
\providecommand \Eprint [0]{\href }%
\providecommand \doibase [0]{http://dx.doi.org/}%
\providecommand \selectlanguage [0]{\@gobble}%
\providecommand \bibinfo  [0]{\@secondoftwo}%
\providecommand \bibfield  [0]{\@secondoftwo}%
\providecommand \translation [1]{[#1]}%
\providecommand \BibitemOpen [0]{}%
\providecommand \bibitemStop [0]{}%
\providecommand \bibitemNoStop [0]{.\EOS\space}%
\providecommand \EOS [0]{\spacefactor3000\relax}%
\providecommand \BibitemShut  [1]{\csname bibitem#1\endcsname}%
\let\auto@bib@innerbib\@empty
%</preamble>
\bibitem [{\citenamefont {Degen}\ \emph {et~al.}(2017)\citenamefont {Degen},
  \citenamefont {Reinhard},\ and\ \citenamefont {Cappellaro}}]{Degen2017}%
  \BibitemOpen
  \bibfield  {author} {\bibinfo {author} {\bibfnamefont {C.~L.}\ \bibnamefont
  {Degen}}, \bibinfo {author} {\bibfnamefont {F.}~\bibnamefont {Reinhard}}, \
  and\ \bibinfo {author} {\bibfnamefont {P.}~\bibnamefont {Cappellaro}},\
  }\href {\doibase 10.1103/RevModPhys.89.035002} {\bibfield  {journal}
  {\bibinfo  {journal} {Rev. Mod. Phys.}\ }\textbf {\bibinfo {volume} {89}},\
  \bibinfo {pages} {035002} (\bibinfo {year} {2017})}\BibitemShut {NoStop}%
\bibitem [{\citenamefont {Mamin}\ and\ \citenamefont
  {Rugar}(2001)}]{Mamin2001}%
  \BibitemOpen
  \bibfield  {author} {\bibinfo {author} {\bibfnamefont {H.~J.}\ \bibnamefont
  {Mamin}}\ and\ \bibinfo {author} {\bibfnamefont {D.}~\bibnamefont {Rugar}},\
  }\href {\doibase 10.1063/1.1418256} {\bibfield  {journal} {\bibinfo
  {journal} {Appl. Phys. Lett.}\ }\textbf {\bibinfo {volume} {79}},\ \bibinfo
  {pages} {3358} (\bibinfo {year} {2001})}\BibitemShut {NoStop}%
\bibitem [{\citenamefont {Teufel}\ \emph {et~al.}(2008)\citenamefont {Teufel},
  \citenamefont {Harlow}, \citenamefont {Regal},\ and\ \citenamefont
  {Lehnert}}]{Teufel2008}%
  \BibitemOpen
  \bibfield  {author} {\bibinfo {author} {\bibfnamefont {J.~D.}\ \bibnamefont
  {Teufel}}, \bibinfo {author} {\bibfnamefont {J.~W.}\ \bibnamefont {Harlow}},
  \bibinfo {author} {\bibfnamefont {C.~A.}\ \bibnamefont {Regal}}, \ and\
  \bibinfo {author} {\bibfnamefont {K.~W.}\ \bibnamefont {Lehnert}},\
  }\href@noop {} {\bibfield  {journal} {\bibinfo  {journal} {Phys. Rev. Lett.}\
  }\textbf {\bibinfo {volume} {101}},\ \bibinfo {pages} {197203} (\bibinfo
  {year} {2008})}\BibitemShut {NoStop}%
\bibitem [{\citenamefont {Jensen}\ \emph {et~al.}(2008)\citenamefont {Jensen},
  \citenamefont {Kim},\ and\ \citenamefont {Zettl}}]{Jensen2008}%
  \BibitemOpen
  \bibfield  {author} {\bibinfo {author} {\bibfnamefont {K.}~\bibnamefont
  {Jensen}}, \bibinfo {author} {\bibfnamefont {K.}~\bibnamefont {Kim}}, \ and\
  \bibinfo {author} {\bibfnamefont {A.}~\bibnamefont {Zettl}},\ }\href
  {\doibase 10.1038/nnano.2008.200} {\bibfield  {journal} {\bibinfo  {journal}
  {Nat. Nanotechnol.}\ }\textbf {\bibinfo {volume} {3}},\ \bibinfo {pages}
  {533} (\bibinfo {year} {2008})}\BibitemShut {NoStop}%
\bibitem [{\citenamefont {Anetsberger}\ \emph {et~al.}(2010)\citenamefont
  {Anetsberger}, \citenamefont {Gavartin}, \citenamefont {Arcizet},
  \citenamefont {Unterreithmeier}, \citenamefont {Weig}, \citenamefont
  {Gorodetsky}, \citenamefont {Kotthaus},\ and\ \citenamefont
  {Kippenberg}}]{Anetsberger2010}%
  \BibitemOpen
  \bibfield  {author} {\bibinfo {author} {\bibfnamefont {G.}~\bibnamefont
  {Anetsberger}}, \bibinfo {author} {\bibfnamefont {E.}~\bibnamefont
  {Gavartin}}, \bibinfo {author} {\bibfnamefont {O.}~\bibnamefont {Arcizet}},
  \bibinfo {author} {\bibfnamefont {Q.~P.}\ \bibnamefont {Unterreithmeier}},
  \bibinfo {author} {\bibfnamefont {E.~M.}\ \bibnamefont {Weig}}, \bibinfo
  {author} {\bibfnamefont {M.~L.}\ \bibnamefont {Gorodetsky}}, \bibinfo
  {author} {\bibfnamefont {J.~P.}\ \bibnamefont {Kotthaus}}, \ and\ \bibinfo
  {author} {\bibfnamefont {T.~J.}\ \bibnamefont {Kippenberg}},\ }\href
  {\doibase 10.1103/PhysRevA.82.061804} {\bibfield  {journal} {\bibinfo
  {journal} {Phys. Rev. A}\ }\textbf {\bibinfo {volume} {82}},\ \bibinfo
  {pages} {061804} (\bibinfo {year} {2010})}\BibitemShut {NoStop}%
\bibitem [{\citenamefont {Moser}\ \emph {et~al.}(2013)\citenamefont {Moser},
  \citenamefont {G{\"{u}}ttinger}, \citenamefont {Eichler}, \citenamefont
  {Esplandiu}, \citenamefont {Liu}, \citenamefont {Dykman},\ and\ \citenamefont
  {Bachtold}}]{Moser2013}%
  \BibitemOpen
  \bibfield  {author} {\bibinfo {author} {\bibfnamefont {J.}~\bibnamefont
  {Moser}}, \bibinfo {author} {\bibfnamefont {J.}~\bibnamefont
  {G{\"{u}}ttinger}}, \bibinfo {author} {\bibfnamefont {A.}~\bibnamefont
  {Eichler}}, \bibinfo {author} {\bibfnamefont {M.~J.}\ \bibnamefont
  {Esplandiu}}, \bibinfo {author} {\bibfnamefont {D.~E.}\ \bibnamefont {Liu}},
  \bibinfo {author} {\bibfnamefont {M.~I.}\ \bibnamefont {Dykman}}, \ and\
  \bibinfo {author} {\bibfnamefont {A.}~\bibnamefont {Bachtold}},\ }\href
  {\doibase 10.1038/nnano.2013.97} {\bibfield  {journal} {\bibinfo  {journal}
  {Nat. Nanotechnol.}\ }\textbf {\bibinfo {volume} {8}},\ \bibinfo {pages}
  {493} (\bibinfo {year} {2013})}\BibitemShut {NoStop}%
\bibitem [{\citenamefont {Chan}\ \emph {et~al.}(2011)\citenamefont {Chan},
  \citenamefont {Alegre}, \citenamefont {Safavi-Naeini}, \citenamefont {Hill},
  \citenamefont {Krause}, \citenamefont {Gr{\"{o}}blacher}, \citenamefont
  {Aspelmeyer},\ and\ \citenamefont {Painter}}]{Chan2011}%
  \BibitemOpen
  \bibfield  {author} {\bibinfo {author} {\bibfnamefont {J.}~\bibnamefont
  {Chan}}, \bibinfo {author} {\bibfnamefont {T.~P.~M.}\ \bibnamefont {Alegre}},
  \bibinfo {author} {\bibfnamefont {A.~H.}\ \bibnamefont {Safavi-Naeini}},
  \bibinfo {author} {\bibfnamefont {J.~T.}\ \bibnamefont {Hill}}, \bibinfo
  {author} {\bibfnamefont {A.}~\bibnamefont {Krause}}, \bibinfo {author}
  {\bibfnamefont {S.}~\bibnamefont {Gr{\"{o}}blacher}}, \bibinfo {author}
  {\bibfnamefont {M.}~\bibnamefont {Aspelmeyer}}, \ and\ \bibinfo {author}
  {\bibfnamefont {O.}~\bibnamefont {Painter}},\ }\href {\doibase
  10.1038/nature10461} {\bibfield  {journal} {\bibinfo  {journal} {Nature}\
  }\textbf {\bibinfo {volume} {478}},\ \bibinfo {pages} {89} (\bibinfo {year}
  {2011})}\BibitemShut {NoStop}%
\bibitem [{\citenamefont {Li}\ \emph {et~al.}(2011)\citenamefont {Li},
  \citenamefont {Kheifets},\ and\ \citenamefont {Raizen}}]{Li2011}%
  \BibitemOpen
  \bibfield  {author} {\bibinfo {author} {\bibfnamefont {T.}~\bibnamefont
  {Li}}, \bibinfo {author} {\bibfnamefont {S.}~\bibnamefont {Kheifets}}, \ and\
  \bibinfo {author} {\bibfnamefont {M.~G.}\ \bibnamefont {Raizen}},\ }\href
  {\doibase 10.1038/nphys1952} {\bibfield  {journal} {\bibinfo  {journal} {Nat.
  Phys.}\ }\textbf {\bibinfo {volume} {7}},\ \bibinfo {pages} {527} (\bibinfo
  {year} {2011})}\BibitemShut {NoStop}%
\bibitem [{\citenamefont {Gieseler}\ \emph {et~al.}(2012)\citenamefont
  {Gieseler}, \citenamefont {Deutsch}, \citenamefont {Quidant},\ and\
  \citenamefont {Novotny}}]{Gieseler2012}%
  \BibitemOpen
  \bibfield  {author} {\bibinfo {author} {\bibfnamefont {J.}~\bibnamefont
  {Gieseler}}, \bibinfo {author} {\bibfnamefont {B.}~\bibnamefont {Deutsch}},
  \bibinfo {author} {\bibfnamefont {R.}~\bibnamefont {Quidant}}, \ and\
  \bibinfo {author} {\bibfnamefont {L.}~\bibnamefont {Novotny}},\ }\href
  {\doibase 10.1103/PhysRevLett.109.103603} {\bibfield  {journal} {\bibinfo
  {journal} {Phys. Rev. Lett.}\ }\textbf {\bibinfo {volume} {109}},\ \bibinfo
  {pages} {103603} (\bibinfo {year} {2012})}\BibitemShut {NoStop}%
\bibitem [{\citenamefont {Chang}\ \emph {et~al.}(2010)\citenamefont {Chang},
  \citenamefont {Regal}, \citenamefont {Papp}, \citenamefont {Wilson},
  \citenamefont {Ye}, \citenamefont {Painter}, \citenamefont {Kimble},\ and\
  \citenamefont {Zoller}}]{Chang2010}%
  \BibitemOpen
  \bibfield  {author} {\bibinfo {author} {\bibfnamefont {D.~E.}\ \bibnamefont
  {Chang}}, \bibinfo {author} {\bibfnamefont {C.~A.}\ \bibnamefont {Regal}},
  \bibinfo {author} {\bibfnamefont {S.~B.}\ \bibnamefont {Papp}}, \bibinfo
  {author} {\bibfnamefont {D.~J.}\ \bibnamefont {Wilson}}, \bibinfo {author}
  {\bibfnamefont {J.}~\bibnamefont {Ye}}, \bibinfo {author} {\bibfnamefont
  {O.}~\bibnamefont {Painter}}, \bibinfo {author} {\bibfnamefont {H.~J.}\
  \bibnamefont {Kimble}}, \ and\ \bibinfo {author} {\bibfnamefont
  {P.}~\bibnamefont {Zoller}},\ }\href {\doibase 10.1073/pnas.0912969107}
  {\bibfield  {journal} {\bibinfo  {journal} {Proc. Natl. Acad. Sci. U.S.A.}\
  }\textbf {\bibinfo {volume} {107}},\ \bibinfo {pages} {1005} (\bibinfo {year}
  {2010})}\BibitemShut {NoStop}%
\bibitem [{\citenamefont {Romero-Isart}\ \emph {et~al.}(2011)\citenamefont
  {Romero-Isart}, \citenamefont {Pflanzer}, \citenamefont {Blaser},
  \citenamefont {Kaltenbaek}, \citenamefont {Kiesel}, \citenamefont
  {Aspelmeyer},\ and\ \citenamefont {Cirac}}]{Romero-Isart2011}%
  \BibitemOpen
  \bibfield  {author} {\bibinfo {author} {\bibfnamefont {O.}~\bibnamefont
  {Romero-Isart}}, \bibinfo {author} {\bibfnamefont {A.~C.}\ \bibnamefont
  {Pflanzer}}, \bibinfo {author} {\bibfnamefont {F.}~\bibnamefont {Blaser}},
  \bibinfo {author} {\bibfnamefont {R.}~\bibnamefont {Kaltenbaek}}, \bibinfo
  {author} {\bibfnamefont {N.}~\bibnamefont {Kiesel}}, \bibinfo {author}
  {\bibfnamefont {M.}~\bibnamefont {Aspelmeyer}}, \ and\ \bibinfo {author}
  {\bibfnamefont {J.~I.}\ \bibnamefont {Cirac}},\ }\href {\doibase
  10.1103/PhysRevLett.107.020405} {\bibfield  {journal} {\bibinfo  {journal}
  {Phys. Rev. Lett.}\ }\textbf {\bibinfo {volume} {107}},\ \bibinfo {pages}
  {020405} (\bibinfo {year} {2011})}\BibitemShut {NoStop}%
\bibitem [{\citenamefont {Arndt}\ and\ \citenamefont
  {Hornberger}(2014)}]{Arndt2014}%
  \BibitemOpen
  \bibfield  {author} {\bibinfo {author} {\bibfnamefont {M.}~\bibnamefont
  {Arndt}}\ and\ \bibinfo {author} {\bibfnamefont {K.}~\bibnamefont
  {Hornberger}},\ }\href {\doibase 10.1038/nphys2863} {\bibfield  {journal}
  {\bibinfo  {journal} {Nat. Phys.}\ }\textbf {\bibinfo {volume} {10}},\
  \bibinfo {pages} {271} (\bibinfo {year} {2014})}\BibitemShut {NoStop}%
\bibitem [{\citenamefont {Jain}\ \emph {et~al.}(2016)\citenamefont {Jain},
  \citenamefont {Gieseler}, \citenamefont {Moritz}, \citenamefont {Dellago},
  \citenamefont {Quidant},\ and\ \citenamefont {Novotny}}]{Jain2016}%
  \BibitemOpen
  \bibfield  {author} {\bibinfo {author} {\bibfnamefont {V.}~\bibnamefont
  {Jain}}, \bibinfo {author} {\bibfnamefont {J.}~\bibnamefont {Gieseler}},
  \bibinfo {author} {\bibfnamefont {C.}~\bibnamefont {Moritz}}, \bibinfo
  {author} {\bibfnamefont {C.}~\bibnamefont {Dellago}}, \bibinfo {author}
  {\bibfnamefont {R.}~\bibnamefont {Quidant}}, \ and\ \bibinfo {author}
  {\bibfnamefont {L.}~\bibnamefont {Novotny}},\ }\href {\doibase
  10.1103/PhysRevLett.116.243601} {\bibfield  {journal} {\bibinfo  {journal}
  {Phys. Rev. Lett.}\ }\textbf {\bibinfo {volume} {116}},\ \bibinfo {pages}
  {243601} (\bibinfo {year} {2016})}\BibitemShut {NoStop}%
\bibitem [{\citenamefont {Kiesel}\ \emph {et~al.}(2013)\citenamefont {Kiesel},
  \citenamefont {Blaser}, \citenamefont {Delic}, \citenamefont {Grass},
  \citenamefont {Kaltenbaek},\ and\ \citenamefont {Aspelmeyer}}]{Kiesel2013a}%
  \BibitemOpen
  \bibfield  {author} {\bibinfo {author} {\bibfnamefont {N.}~\bibnamefont
  {Kiesel}}, \bibinfo {author} {\bibfnamefont {F.}~\bibnamefont {Blaser}},
  \bibinfo {author} {\bibfnamefont {U.}~\bibnamefont {Delic}}, \bibinfo
  {author} {\bibfnamefont {D.}~\bibnamefont {Grass}}, \bibinfo {author}
  {\bibfnamefont {R.}~\bibnamefont {Kaltenbaek}}, \ and\ \bibinfo {author}
  {\bibfnamefont {M.}~\bibnamefont {Aspelmeyer}},\ }\href {\doibase
  10.1073/pnas.1309167110} {\bibfield  {journal} {\bibinfo  {journal} {Proc.
  Natl. Acad. Sci. U.S.A.}\ }\textbf {\bibinfo {volume} {110}},\ \bibinfo
  {pages} {14180} (\bibinfo {year} {2013})}\BibitemShut {NoStop}%
\bibitem [{\citenamefont {Asenbaum}\ \emph {et~al.}(2013)\citenamefont
  {Asenbaum}, \citenamefont {Kuhn}, \citenamefont {Nimmrichter}, \citenamefont
  {Sezer},\ and\ \citenamefont {Arndt}}]{Asenbaum2013}%
  \BibitemOpen
  \bibfield  {author} {\bibinfo {author} {\bibfnamefont {P.}~\bibnamefont
  {Asenbaum}}, \bibinfo {author} {\bibfnamefont {S.}~\bibnamefont {Kuhn}},
  \bibinfo {author} {\bibfnamefont {S.}~\bibnamefont {Nimmrichter}}, \bibinfo
  {author} {\bibfnamefont {U.}~\bibnamefont {Sezer}}, \ and\ \bibinfo {author}
  {\bibfnamefont {M.}~\bibnamefont {Arndt}},\ }\href {\doibase
  10.1038/ncomms3743} {\bibfield  {journal} {\bibinfo  {journal} {Nat.
  Commun.}\ }\textbf {\bibinfo {volume} {4}},\ \bibinfo {pages} {2743}
  (\bibinfo {year} {2013})}\BibitemShut {NoStop}%
\bibitem [{\citenamefont {Millen}\ \emph {et~al.}(2015)\citenamefont {Millen},
  \citenamefont {Fonseca}, \citenamefont {Mavrogordatos}, \citenamefont
  {Monteiro},\ and\ \citenamefont {Barker}}]{Millen2015}%
  \BibitemOpen
  \bibfield  {author} {\bibinfo {author} {\bibfnamefont {J.}~\bibnamefont
  {Millen}}, \bibinfo {author} {\bibfnamefont {P.~Z.~G.}\ \bibnamefont
  {Fonseca}}, \bibinfo {author} {\bibfnamefont {T.}~\bibnamefont
  {Mavrogordatos}}, \bibinfo {author} {\bibfnamefont {T.~S.}\ \bibnamefont
  {Monteiro}}, \ and\ \bibinfo {author} {\bibfnamefont {P.~F.}\ \bibnamefont
  {Barker}},\ }\href {\doibase 10.1103/PhysRevLett.114.123602} {\bibfield
  {journal} {\bibinfo  {journal} {Phys. Rev. Lett.}\ }\textbf {\bibinfo
  {volume} {114}},\ \bibinfo {pages} {123602} (\bibinfo {year}
  {2015})}\BibitemShut {NoStop}%
\bibitem [{\citenamefont {Hackerm{\"{u}}ller}\ \emph
  {et~al.}(2004)\citenamefont {Hackerm{\"{u}}ller}, \citenamefont {Hornberger},
  \citenamefont {Brezger}, \citenamefont {Zeilinger},\ and\ \citenamefont
  {Arndt}}]{Hackermuller2004}%
  \BibitemOpen
  \bibfield  {author} {\bibinfo {author} {\bibfnamefont {L.}~\bibnamefont
  {Hackerm{\"{u}}ller}}, \bibinfo {author} {\bibfnamefont {K.}~\bibnamefont
  {Hornberger}}, \bibinfo {author} {\bibfnamefont {B.}~\bibnamefont {Brezger}},
  \bibinfo {author} {\bibfnamefont {A.}~\bibnamefont {Zeilinger}}, \ and\
  \bibinfo {author} {\bibfnamefont {M.}~\bibnamefont {Arndt}},\ }\href
  {\doibase 10.1038/nature02276} {\bibfield  {journal} {\bibinfo  {journal}
  {Nature}\ }\textbf {\bibinfo {volume} {427}},\ \bibinfo {pages} {711}
  (\bibinfo {year} {2004})}\BibitemShut {NoStop}%
\bibitem [{\citenamefont {Bateman}\ \emph {et~al.}(2014)\citenamefont
  {Bateman}, \citenamefont {Nimmrichter}, \citenamefont {Hornberger},\ and\
  \citenamefont {Ulbricht}}]{Bateman2014}%
  \BibitemOpen
  \bibfield  {author} {\bibinfo {author} {\bibfnamefont {J.}~\bibnamefont
  {Bateman}}, \bibinfo {author} {\bibfnamefont {S.}~\bibnamefont
  {Nimmrichter}}, \bibinfo {author} {\bibfnamefont {K.}~\bibnamefont
  {Hornberger}}, \ and\ \bibinfo {author} {\bibfnamefont {H.}~\bibnamefont
  {Ulbricht}},\ }\href {\doibase 10.1038/ncomms5788} {\bibfield  {journal}
  {\bibinfo  {journal} {Nat. Commun.}\ }\textbf {\bibinfo {volume} {5}},\
  \bibinfo {pages} {4788} (\bibinfo {year} {2014})}\BibitemShut {NoStop}%
\bibitem [{\citenamefont {Millen}\ \emph {et~al.}(2014)\citenamefont {Millen},
  \citenamefont {Deesuwan}, \citenamefont {Barker},\ and\ \citenamefont
  {Anders}}]{Millen2014}%
  \BibitemOpen
  \bibfield  {author} {\bibinfo {author} {\bibfnamefont {J.}~\bibnamefont
  {Millen}}, \bibinfo {author} {\bibfnamefont {T.}~\bibnamefont {Deesuwan}},
  \bibinfo {author} {\bibfnamefont {P.}~\bibnamefont {Barker}}, \ and\ \bibinfo
  {author} {\bibfnamefont {J.}~\bibnamefont {Anders}},\ }\href {\doibase
  10.1038/nnano.2014.82} {\bibfield  {journal} {\bibinfo  {journal} {Nat.
  Nanotechnol.}\ }\textbf {\bibinfo {volume} {9}},\ \bibinfo {pages} {425}
  (\bibinfo {year} {2014})}\BibitemShut {NoStop}%
\bibitem [{\citenamefont {Ranjit}\ \emph {et~al.}(2015)\citenamefont {Ranjit},
  \citenamefont {Atherton}, \citenamefont {Stutz}, \citenamefont {Cunningham},\
  and\ \citenamefont {Geraci}}]{Ranjit2015a}%
  \BibitemOpen
  \bibfield  {author} {\bibinfo {author} {\bibfnamefont {G.}~\bibnamefont
  {Ranjit}}, \bibinfo {author} {\bibfnamefont {D.~P.}\ \bibnamefont
  {Atherton}}, \bibinfo {author} {\bibfnamefont {J.~H.}\ \bibnamefont {Stutz}},
  \bibinfo {author} {\bibfnamefont {M.}~\bibnamefont {Cunningham}}, \ and\
  \bibinfo {author} {\bibfnamefont {A.~A.}\ \bibnamefont {Geraci}},\ }\href
  {\doibase 10.1103/PhysRevA.91.051805} {\bibfield  {journal} {\bibinfo
  {journal} {Phys. Rev. A}\ }\textbf {\bibinfo {volume} {91}},\ \bibinfo
  {pages} {051805} (\bibinfo {year} {2015})}\BibitemShut {NoStop}%
\bibitem [{\citenamefont {Rahman}\ \emph {et~al.}(2016)\citenamefont {Rahman},
  \citenamefont {Frangeskou}, \citenamefont {Kim}, \citenamefont {Bose},
  \citenamefont {Morley},\ and\ \citenamefont {Barker}}]{Rahman2015a}%
  \BibitemOpen
  \bibfield  {author} {\bibinfo {author} {\bibfnamefont {A.~T. M.~A.}\
  \bibnamefont {Rahman}}, \bibinfo {author} {\bibfnamefont {A.~C.}\
  \bibnamefont {Frangeskou}}, \bibinfo {author} {\bibfnamefont {M.~S.}\
  \bibnamefont {Kim}}, \bibinfo {author} {\bibfnamefont {S.}~\bibnamefont
  {Bose}}, \bibinfo {author} {\bibfnamefont {G.~W.}\ \bibnamefont {Morley}}, \
  and\ \bibinfo {author} {\bibfnamefont {P.~F.}\ \bibnamefont {Barker}},\
  }\href {\doibase 10.1038/srep21633} {\bibfield  {journal} {\bibinfo
  {journal} {Sci. Rep.}\ }\textbf {\bibinfo {volume} {6}},\ \bibinfo {pages}
  {21633} (\bibinfo {year} {2016})}\BibitemShut {NoStop}%
\bibitem [{\citenamefont {Juan}\ \emph {et~al.}(2016)\citenamefont {Juan},
  \citenamefont {Molina-Terriza}, \citenamefont {Volz},\ and\ \citenamefont
  {Romero-Isart}}]{Juan2016}%
  \BibitemOpen
  \bibfield  {author} {\bibinfo {author} {\bibfnamefont {M.~L.}\ \bibnamefont
  {Juan}}, \bibinfo {author} {\bibfnamefont {G.}~\bibnamefont
  {Molina-Terriza}}, \bibinfo {author} {\bibfnamefont {T.}~\bibnamefont
  {Volz}}, \ and\ \bibinfo {author} {\bibfnamefont {O.}~\bibnamefont
  {Romero-Isart}},\ }\href {\doibase 10.1103/PhysRevA.94.023841} {\bibfield
  {journal} {\bibinfo  {journal} {Phys. Rev. A}\ }\textbf {\bibinfo {volume}
  {94}},\ \bibinfo {pages} {023841} (\bibinfo {year} {2016})}\BibitemShut
  {NoStop}%
\bibitem [{\citenamefont {Rahman}\ and\ \citenamefont
  {Barker}(2017)}]{Rahman2017}%
  \BibitemOpen
  \bibfield  {author} {\bibinfo {author} {\bibfnamefont {A.~T. M.~A.}\
  \bibnamefont {Rahman}}\ and\ \bibinfo {author} {\bibfnamefont {P.~F.}\
  \bibnamefont {Barker}},\ }\href {\doibase 10.1038/s41566-017-0005-3}
  {\bibfield  {journal} {\bibinfo  {journal} {Nat. Photonics}\ }\textbf
  {\bibinfo {volume} {11}},\ \bibinfo {pages} {634} (\bibinfo {year}
  {2017})}\BibitemShut {NoStop}%
\bibitem [{\citenamefont {Ranjit}\ \emph {et~al.}(2016)\citenamefont {Ranjit},
  \citenamefont {Cunningham}, \citenamefont {Casey},\ and\ \citenamefont
  {Geraci}}]{Ranjit2016}%
  \BibitemOpen
  \bibfield  {author} {\bibinfo {author} {\bibfnamefont {G.}~\bibnamefont
  {Ranjit}}, \bibinfo {author} {\bibfnamefont {M.}~\bibnamefont {Cunningham}},
  \bibinfo {author} {\bibfnamefont {K.}~\bibnamefont {Casey}}, \ and\ \bibinfo
  {author} {\bibfnamefont {A.~A.}\ \bibnamefont {Geraci}},\ }\href {\doibase
  10.1103/PhysRevA.93.053801} {\bibfield  {journal} {\bibinfo  {journal} {Phys.
  Rev. A}\ }\textbf {\bibinfo {volume} {93}},\ \bibinfo {pages} {053801}
  (\bibinfo {year} {2016})}\BibitemShut {NoStop}%
\bibitem [{\citenamefont {Hebestreit}\ \emph {et~al.}(2017)\citenamefont
  {Hebestreit}, \citenamefont {Frimmer}, \citenamefont {Reimann}, \citenamefont
  {Dellago}, \citenamefont {Ricci},\ and\ \citenamefont
  {Novotny}}]{Hebestreit2017}%
  \BibitemOpen
  \bibfield  {author} {\bibinfo {author} {\bibfnamefont {E.}~\bibnamefont
  {Hebestreit}}, \bibinfo {author} {\bibfnamefont {M.}~\bibnamefont {Frimmer}},
  \bibinfo {author} {\bibfnamefont {R.}~\bibnamefont {Reimann}}, \bibinfo
  {author} {\bibfnamefont {C.}~\bibnamefont {Dellago}}, \bibinfo {author}
  {\bibfnamefont {F.}~\bibnamefont {Ricci}}, \ and\ \bibinfo {author}
  {\bibfnamefont {L.}~\bibnamefont {Novotny}},\ }\href@noop {} {\  (\bibinfo
  {year} {2017})},\ \Eprint {http://arxiv.org/abs/1711.09049}
  {arXiv:1711.09049} \BibitemShut {NoStop}%
\bibitem [{\citenamefont {Kitamura}\ \emph {et~al.}(2007)\citenamefont
  {Kitamura}, \citenamefont {Pilon},\ and\ \citenamefont
  {Jonasz}}]{Kitamura2007}%
  \BibitemOpen
  \bibfield  {author} {\bibinfo {author} {\bibfnamefont {R.}~\bibnamefont
  {Kitamura}}, \bibinfo {author} {\bibfnamefont {L.}~\bibnamefont {Pilon}}, \
  and\ \bibinfo {author} {\bibfnamefont {M.}~\bibnamefont {Jonasz}},\ }\href
  {\doibase 10.1364/AO.46.008118} {\bibfield  {journal} {\bibinfo  {journal}
  {Appl. Opt.}\ }\textbf {\bibinfo {volume} {46}},\ \bibinfo {pages} {8118}
  (\bibinfo {year} {2007})}\BibitemShut {NoStop}%
\bibitem [{\citenamefont {Gieseler}\ \emph {et~al.}(2014)\citenamefont
  {Gieseler}, \citenamefont {Quidant}, \citenamefont {Dellago},\ and\
  \citenamefont {Novotny}}]{Gieseler2014a}%
  \BibitemOpen
  \bibfield  {author} {\bibinfo {author} {\bibfnamefont {J.}~\bibnamefont
  {Gieseler}}, \bibinfo {author} {\bibfnamefont {R.}~\bibnamefont {Quidant}},
  \bibinfo {author} {\bibfnamefont {C.}~\bibnamefont {Dellago}}, \ and\
  \bibinfo {author} {\bibfnamefont {L.}~\bibnamefont {Novotny}},\ }\href
  {\doibase 10.1038/nnano.2014.40} {\bibfield  {journal} {\bibinfo  {journal}
  {Nat. Nanotechnol.}\ }\textbf {\bibinfo {volume} {9}},\ \bibinfo {pages}
  {358} (\bibinfo {year} {2014})}\BibitemShut {NoStop}%
\bibitem [{\citenamefont {Br{\"{u}}ckner}(1970)}]{Brueckner1970}%
  \BibitemOpen
  \bibfield  {author} {\bibinfo {author} {\bibfnamefont {R.}~\bibnamefont
  {Br{\"{u}}ckner}},\ }\href {\doibase 10.1016/0022-3093(70)90190-0} {\bibfield
   {journal} {\bibinfo  {journal} {J. Non-Cryst. Solids}\ }\textbf {\bibinfo
  {volume} {5}},\ \bibinfo {pages} {123} (\bibinfo {year} {1970})}\BibitemShut
  {NoStop}%
\bibitem [{\citenamefont {Waxler}\ and\ \citenamefont
  {Cleek}(1973)}]{Waxler1973}%
  \BibitemOpen
  \bibfield  {author} {\bibinfo {author} {\bibfnamefont {R.~M.}\ \bibnamefont
  {Waxler}}\ and\ \bibinfo {author} {\bibfnamefont {G.}~\bibnamefont {Cleek}},\
  }\href {\doibase 10.6028/jres.077A.046} {\bibfield  {journal} {\bibinfo
  {journal} {J. Res. Natl. Stand. Sec. A}\ }\textbf {\bibinfo {volume} {77A}},\
  \bibinfo {pages} {755} (\bibinfo {year} {1973})}\BibitemShut {NoStop}%
\bibitem [{\citenamefont {Epstein}(1924)}]{Epstein1924}%
  \BibitemOpen
  \bibfield  {author} {\bibinfo {author} {\bibfnamefont {P.}~\bibnamefont
  {Epstein}},\ }\href {\doibase 10.1103/PhysRev.23.710} {\bibfield  {journal}
  {\bibinfo  {journal} {Phys. Rev.}\ }\textbf {\bibinfo {volume} {23}},\
  \bibinfo {pages} {710} (\bibinfo {year} {1924})}\BibitemShut {NoStop}%
\bibitem [{\citenamefont {Kaltenbaek}\ \emph {et~al.}(2012)\citenamefont
  {Kaltenbaek}, \citenamefont {Hechenblaikner}, \citenamefont {Kiesel},
  \citenamefont {Romero-Isart}, \citenamefont {Schwab}, \citenamefont
  {Johann},\ and\ \citenamefont {Aspelmeyer}}]{Kaltenbaek2012}%
  \BibitemOpen
  \bibfield  {author} {\bibinfo {author} {\bibfnamefont {R.}~\bibnamefont
  {Kaltenbaek}}, \bibinfo {author} {\bibfnamefont {G.}~\bibnamefont
  {Hechenblaikner}}, \bibinfo {author} {\bibfnamefont {N.}~\bibnamefont
  {Kiesel}}, \bibinfo {author} {\bibfnamefont {O.}~\bibnamefont
  {Romero-Isart}}, \bibinfo {author} {\bibfnamefont {K.~C.}\ \bibnamefont
  {Schwab}}, \bibinfo {author} {\bibfnamefont {U.}~\bibnamefont {Johann}}, \
  and\ \bibinfo {author} {\bibfnamefont {M.}~\bibnamefont {Aspelmeyer}},\
  }\href {\doibase 10.1007/s10686-012-9292-3} {\bibfield  {journal} {\bibinfo
  {journal} {Exp. Astron.}\ }\textbf {\bibinfo {volume} {34}},\ \bibinfo
  {pages} {123} (\bibinfo {year} {2012})}\BibitemShut {NoStop}%
\bibitem [{\citenamefont {Romero-Isart}\ \emph {et~al.}(2012)\citenamefont
  {Romero-Isart}, \citenamefont {Clemente}, \citenamefont {Navau},
  \citenamefont {Sanchez},\ and\ \citenamefont {Cirac}}]{Romero-Isart2012}%
  \BibitemOpen
  \bibfield  {author} {\bibinfo {author} {\bibfnamefont {O.}~\bibnamefont
  {Romero-Isart}}, \bibinfo {author} {\bibfnamefont {L.}~\bibnamefont
  {Clemente}}, \bibinfo {author} {\bibfnamefont {C.}~\bibnamefont {Navau}},
  \bibinfo {author} {\bibfnamefont {A.}~\bibnamefont {Sanchez}}, \ and\
  \bibinfo {author} {\bibfnamefont {J.~I.}\ \bibnamefont {Cirac}},\ }\href
  {\doibase 10.1103/PhysRevLett.109.147205} {\bibfield  {journal} {\bibinfo
  {journal} {Phys. Rev. Lett.}\ }\textbf {\bibinfo {volume} {109}},\ \bibinfo
  {pages} {147205} (\bibinfo {year} {2012})}\BibitemShut {NoStop}%
\end{thebibliography}
%

\end{document}